\documentclass[12pt]{article}
\usepackage{amsfonts,graphicx,amssymb,amsmath,amsthm,epsfig}
\usepackage{float}
\allowdisplaybreaks

\begin{document}
\title{\bf Study of Charged Celestial Objects in Modified Gravity}
\author{M. Sharif$^1$ \thanks{msharif.math@pu.edu.pk}
and K. Hassan$^2$ \thanks{komalhassan3@gmail.com}\\
$^1$ Department of Mathematics and Statistics, The University of Lahore,\\
1-KM Defence Road Lahore, Pakistan.\\
$^2$ Department of Mathematics, University of the Punjab,\\
Quaid-e-Azam Campus, Lahore-54590, Pakistan.}
\date{}

\maketitle
\begin{abstract}
In this paper, we assess different charged self-gravitating stellar
models possessing anisotropic matter source in the background of
$f(G,T)$ gravity. For this purpose, we choose a well-known model of
this gravity, i.e., $f(G,T)=G^2+\varrho T$, where $\varrho$ stands
for the coupling constant. The modified field equations are
developed using MIT bag model equation of state, and their solution
is found with the help of Tolman IV ansatz which contains three
unknown constants. This solution is further exploited to examine the
graphical behavior of Her X-I, PSR J1614-2230, 4U1820-30 and LMC X-4
celestial objects. We assume two different values of charge to
figure out the pressure constituents, energy density, anisotropy and
energy constraints graphically. We also discuss compactness, mass
and redshift parameters. Finally, we explore stability of the
considered stars through two different methods. It is concluded that
all the star candidates are viable as well as stable for
$\mathcal{Q}=0.1$. For the larger charge, the viable behavior is
also observed for all stars but PSR J1614-2230 shows unstable trend.
\end{abstract}

\section{Introduction}

The eye-catching fundamental astrophysical objects, i.e., stars are
present alongside other constituents in the cosmos. Astrophysicists
got captivated by these objects and devoted their attention to
studying the secrets behind their evolutionary processes. The
existence of heat and luminosity in the cosmic world is the result
of multiple nuclear reactions that occur in the middle of celestial
bodies. In the meantime, there arises a stage in which the
gravitational force (acting internally) becomes dominant over the
outward-producing pressure. It leads to the collapse of astronomical
bodies and the beginning of some new celestial entities, namely
white dwarf, neutron star, and black hole. In order to stop further
collapse in neutron stars, the neutrons produce degeneracy pressure,
which operates in the opposite direction of gravity. The
anticipation of neutron stars was predicted for the first time in
1934 \cite{1}, although their emergence was confirmed afterwards. A
strange star is an extremely dense intermediate state that exists
between neutron stars and black holes and includes quark matter.
These stars are thought to have interiors made up of up, down, and
strange quarks, which are far more massive than neutron stars. Many
researchers made early attempts to comprehend these hypothetical
systems \cite{2}.

The anisotropic matter distribution is assumed to be a good source
of study to understand the interior composition and dynamics of
stellar objects. Pressure anisotropy is caused by the discrepancy
between radial and tangential pressures. Ruderman \cite{3} suggested
that anisotropy in the compact bodies with denser cores is generated
due to the interacting nuclear matter. The physical attributes of
anisotropic stellar structures are studied by Dev and Gleiser
\cite{5} using various equations of state (EoS) that relate
tangential and radial pressure components. Hossein et al \cite{6}
investigated the key characteristics of 4U1820-30 through graphical
interpretation by including the cosmological constant in the field
equations. Kalam et al \cite{7} checked the viability and stability
of numerous neutron stars. The embedding class 1 solution is used to
derive the anisotropic results through which the physical acceptance
conditions of stars are checked by Maurya et al \cite{8}. The
inclusion of electric charge in the inner distribution provides a
comprehensive way to apprehend the expansion and stability of
celestial structures. Charges produce a force in the outward
direction that acts against gravity and supports the objects in
sustaining their existence for a longer time. Murad \cite{8a}
assumed the anisotropic effects on charged stellar bodies to derive
their solution and visualized them graphically. The condition of
conformal symmetry has been utilized in extracting the effects of
electromagnetic field on realistic models by Matondo et al
\cite{8b}. A significant work to realize the contribution of
electric charge on different structures has been accomplished in
\cite{8c}.

The MIT bag model EoS is found more useful in describing the
interior structure of quark objects. It is important to mention here
that the compactness of the astronomical stars 4U 1820-30, SAX J
1808.4-3658, Her X-1, 4U 1728-34, PSR 0943+10, RXJ 185635- 3754 was
not proficiently measured through neutron star EoS, whereas more
considerable results have been produced with quark matter EoS
\cite{9}. The difference between real and false vacuums can be
evaluated with the help of bag constant present in the bag model
EoS. It is observed that the increment in the values of
$\mathfrak{B}_c$ decreases the quark pressure. For the exploration
of inner configuration of quark stars associated with MIT bag EoS,
there is a large body of literature. When Demorest et al \cite{9a}
observed the candidate PSR J1614-2230, they found that these kinds
of massive bodies can be elucidated more efficiently through MIT bag
EoS. Using the interpolation method, Rahaman et al. \cite{10}
determined the mass of quark star with a radius of 9.9 km and used
this finding to explain several physical aspects. Bhar \cite{11}
explored certain compact entities experiencing anisotropy in the
light of Krori-Barua ansatz and MIT bag EoS. Using corresponding
EoS, Deb et al \cite{12} examined the essential constraints to be
fulfilled for physically acceptable stars in the case of
charged/uncharged sources. Sharif and his collaborators \cite{13}
developed non-singular solution of the field equations to analyze
the anisotropic stellar systems.

The general theory of relativity (GR) is taken into account as a
foundation to comprehend the gravitational interaction at large
scales. This has played a crucial role in opening new windows to the
discovery of mysterious entities. The existence of dark energy,
which has large negative pressure and is thought to be the primary
source of cosmic development, as well as dark matter, are two main
issues that GR does not adequately address. Thus, an attempt to
uncover these mysterious components is achieved by altering the
Einstein-Hilbert action that gives rise to the formulation of
modified gravity.

In this respect, the Ricci scalar $R$ in the Einstein-Hilbert action
is interchanged from its generalized function to yield the first
modified theory namely, $f(R)$ gravity. Various researchers have
inspected inflationary and current cosmic acceleration in the realm
of feasible $f(R)$ models \cite{32}. Also, a large number of stellar
bodies possessing fundamental characteristics have been discussed by
employing different methods in this theory \cite{33}. A wide range
of modified theories with arbitrary coupling are of great interest
to astrophysicists. In order to obtain the exciting results, Harko
et al \cite{34} coupled geometry with matter in the action and
proposed $f(R,T)$ gravity, where $T$ indicates the trace of
energy-momentum tensor. The impact of minimal/non-minimal
interactions on test particles has been analyzed using several
$f(R,T)$ models. Numerous researchers have examined various
configurations and noticed fascinating astrophysical findings in
this theory \cite{35}.

In addition to the aforementioned theories, $f(G)$ gravity is
another modification of GR which is formulated by Nojiri and
Odintsov \cite{36} by adding $f(G)$ ($G$ is the Gauss-Bonnet
invariant (GB)) with $R$. Moreover, the expression of GB is
specified as
$G=R^{\phi\sigma\gamma\delta}R_{\phi\sigma\gamma\delta}+R^2-4R^{\phi\sigma}R_{\phi\sigma}$,
with $R_{\phi\sigma}$ and $R_{\phi\sigma\gamma\delta}$ being the
Ricci and Riemann tensors, respectively. Four different kinds of
future singularities were investigated under the reconstruction of
feasible $f(G)$ model by Bamba et al \cite{14} corresponding to the
phantom or quintessence accelerating epoch. In modified Gauss-Bonnet
gravity, Myrzakulov et al \cite{15} established the cosmic solutions
without including cosmological constant and effectively described
the existence of dark energy and inflationary epoch. The analysis of
accelerating expansion with viable $f(G)$ model is executed by
computing the symmetry generators according to the separation of
variables and power-law type (employing Noether symmetry approach)
\cite{17}. The values of Hubble, jerk, snap and deceleration
parameters have been kept fixed by Bamba et al \cite{16} in order to
look for the regular configured source by utilizing three $f(G)$
ansatzs. Sharif and Ramzan \cite{18} used the radii and masses of
several astronomical bodies to observe their viable and stable
characters under the influence of modified Guass-Bonnet theory.

Sharif and Ikram \cite{37} got inspired from the geometry-matter
coupling and developed another novel theory by substituting the
function $f(G,T)$ in place of $f(G)$ and naming it $f(G,T)$ gravity.
Shamir and Ahmad \cite{19} focused on exploring the physical
attributes of three stellar entities undergoing anisotropy in the
framework of $f(G,T)$ gravity. The compact star Vela X-1 was chosen
by Maurya et al \cite{20} to estimate the viable and stable features
of its inner anisotropic source. Sharif and Naeem \cite{21}
discussed the novel characteristics of static anisotropic fluid
spheres in this theory. Recently, we have calculated the $Y_{TF}$
factor to observe the complex structures by employing Herrera's
decomposing technique for static charged and uncharged sources,
non-static cylindrical and spherical (charged/uncharged) systems
\cite{21a}.

In this paper, we focus on revealing the contribution of
electromagnetic field and $f(G,T)$ gravity in studying the physical
aspects of quark candidates. The paper is structured as follows. In
section \textbf{2}, primitive idea of this gravity, crucial
equations and Tolman IV solution are examined. Section \textbf{3}
uses the boundary conditions to generally express the constants (in
Tolman IV) in the observational data of stars. The physical
acceptance and key features of all compact objects corresponding to
two charge values are reviewed in section \textbf{4}. The last
section describes the significant outcomes.

\section{$f(G,T)$ Gravity and Matter Determinants}

The action of $f(G,T)$ gravity together with Ricci scalar in the
Einstein-Maxwell action is given as follows
\begin{equation}\label{1}
\mathfrak{I}_{f(G,T)}=\frac{1}{16\pi}\int
\big(f(G,T)+R\big)d^{4}x\sqrt{-g}+\int\big(\mathfrak{L}_{\mathbf{E}}
+\mathfrak{L}_{\mathbf{M}})d^{4}x\sqrt{-g},
\end{equation}
where $\mathfrak{L}_{\mathbf{E}}$ specifies the Lagrangian density
associated with the electromagnetic field and for the usual matter
distribution it is denoted by $\mathfrak{L}_{\mathbf{M}}$. The
symbol $g$ indicates determinant of the metric tensor and the
Lagrangian density for the present setup is fixed as
$\mathfrak{L}_{\mathbf{M}}=\mathbb{P}$. The relation between the
energy-momentum tensor and Lagrangian density is observed as
\begin{align}\label{1a}
T_{\phi\sigma}=g_{\phi\sigma}\mathfrak{L}_{\mathbf{M}}
-\frac{2\partial\mathfrak{L}_{\mathbf{M}}}{\partial g^{\phi\sigma}}.
\end{align}
In the present system, the action (1) is varied with
$g_{\phi\sigma}$ to develop the field equations which are given as
follows
\begin{eqnarray}\nonumber
G_{\phi\sigma}&=&8\pi
(T^{(\textsf{m})}_{\phi\sigma}+\mathcal{S}_{\phi\sigma})
-(T^{(\textsf{m})}_{\phi\sigma}+\Theta_{\phi\sigma})f_{T}(G,T)
+\frac{1}{2}g_{\phi\sigma}f(G,T) +\big(4R^{\mu\nu}R_{\phi\mu \sigma
\nu}\\\nonumber&-& 2RR_{\phi\sigma}-2R^{\mu \nu \gamma}
_{\phi}R_{\sigma \mu \nu
\gamma}+4R_{\mu\sigma}R^{\mu}_{\phi}\big)f_{G}(G,T)
+(4g_{\phi\sigma}R^{\mu
\nu}\nabla_{\mu}\nabla_{\nu}\\\nonumber&-&4R^{\mu}_{\phi}\nabla_{\sigma}\nabla_{\mu}-4R_{\phi\mu
\sigma\nu}\nabla^{\mu}\nabla^{\nu}-2g_{\phi\sigma}R\nabla^{2}
+2R\nabla_{\phi}\nabla_{\sigma}-4R^{\mu}_{\sigma}\nabla_{\phi}\nabla_{\mu}\\\label{4}
&+&4R_{\phi\sigma}\nabla^{2})f_{G}(G,T),
\end{eqnarray}
where
$\Theta_{\phi\sigma}=-2T^{(\textsf{m})}_{\phi\sigma}+\mathbb{P}g_{\phi\sigma}$
along with $\mathbb{P}=\frac{\mathbb{P}_{r}+2\mathbb{P}_{t}}{3}$ and
the geometrical representation of the celestial bodies is provided
by the Einstein tensor ($G_{\phi\sigma}$). The expressions $f_{T}$
and $f_{G}$ are given as $\frac{\partial f(G,T)}{\partial T}$ and
$\frac{\partial f(G,T)}{\partial G}$, respectively. Moreover, the
term $\nabla_{\phi}$ is the covariant derivative and the d' Alembert
operator is signified by $\nabla^{2}=\nabla^{a}\nabla_{a}=\Box$. The
replacement $f(G,T)=f(G)$ in the action produces the field equations
of $f(G)$ theory. The contribution of extra terms in this theory is
represented by
\begin{align}\nonumber
T^{\textsf{cor}}_{\phi\sigma}&=\frac{1}{8\pi}\bigg[\{(\mathbb{U}
+\mathbb{P}_{t})\textsf{W}_{\phi}\textsf{W}_{\sigma}
-\frac{1}{3}(\mathbb{P}_{r}-\mathbb{P}_{t})g_{\phi\sigma}
+(\mathbb{P}_{r}-\mathbb{P}_{t})\textsf{X}_{\phi}
\textsf{X}_{\sigma}\}f_{T}(G,T)
\\\nonumber&+\frac{1}{2}g_{\phi\sigma}f(G,T)
+\big(4R^{\mu\nu}R_{\phi\mu \sigma \nu}-
2RR_{\phi\sigma}-2R^{\mu \nu \gamma} _{\phi}R_{\sigma \mu
\nu \gamma}+4R_{\mu\sigma}R^{\mu}_{\phi}\big)f_{G}(G,T)
\\\nonumber&+(4g_{\phi\sigma}R^{\mu
\nu}\nabla_{\mu}\nabla_{\nu}-4R^{\mu}_{\phi}\nabla_{\sigma}\nabla_{\mu}-4R_{\phi\mu
\sigma\nu}\nabla^{\mu}\nabla^{\nu}-2g_{\phi\sigma}R\nabla^{2}
+2R\nabla_{\phi}\nabla_{\sigma}\\\label{4}&-4R^{\mu}_{\sigma}\nabla_{\phi}\nabla_{\mu}
+4R_{\phi\sigma}\nabla^{2})f_{G}(G,T)\bigg].
\end{align}

The following energy-momentum tensor is employed to examine the
substantial features of anisotropic stellar candidates as
\begin{equation}\label{5}
T^{(\textsf{m})}_{\phi\sigma} =(\mathbb{U}+\mathbb{P}_{t})
\textsf{W}_{\phi}\textsf{W}_{\sigma}+\mathbb{P}_{t}g_{\phi\sigma}
+(\mathbb{P}_{r}-\mathbb{P}_{t})\textsf{X}_{\phi}\textsf{X}_{\sigma},
\end{equation}
where pressures in the radial and tangential directions are
indicated by $\mathbb{P}_{r}$ and $\mathbb{P}_{t}$, respectively,
and $\mathbb{U}$ stands for the energy density of the fluid source.
The covariant component of four-vector is expressed as
$\textsf{X}_{\phi}$ and of four-velocity is denoted by
$\textsf{W}_{\phi}$, portraying the relations
$\textsf{X}^{\phi}\textsf{W}_{\phi}=0$ and
$\textsf{W}^{\phi}\textsf{W}_{\phi}=-1$. The alternative way of
presenting Eq.\eqref{4} is
\begin{equation}\label{3}
G_{\phi\sigma}=8\pi
T^{\texttt{(tot)}}_{\phi\sigma}=8\pi(T^{\textsf{cor}}_{\phi\sigma}+T^{(\textsf{m})}_{\phi\sigma}).
\end{equation}
The electromagnetic energy tensor to discuss the charged fluid
within the compact bodies is characterized by
\begin{equation}\label{3b}
\mathcal{S}_{\phi\sigma}=\frac{1}{4\pi}\left(\mathcal{F}^{n}_{\phi}\mathcal{F}_{\sigma
n}-\frac{1}{4}\mathcal{F}_{nm}\mathcal{F}^{nm}g_{\phi\sigma}\right),
\end{equation}
where the Maxwell field tensor is specified as
$\mathcal{F}_{\phi\sigma}=\xi_{\sigma,\phi}-\xi_{\phi,\sigma}$ and
the four potential is represented by $\xi_{\phi}$. For the static
symmetric structure, it takes the form
$\xi_{\phi}=\xi(r)\delta^{0}_{\phi}$. The tensorial form of Maxwell
field equations turn out to be
\begin{equation}\nonumber
\mathcal{F}_{[\phi\sigma;n]}=0 ,\quad
\mathcal{F}^{\phi\sigma}_{~~;\sigma}=4\pi \mathcal{J}^{\phi},
\end{equation}
where $\mathcal{J}^{\phi}$ indicates the electromagnetic
four-current vector with
$\mathcal{J}^{\phi}=\vartheta\textsf{W}^{\phi}$ and $\vartheta$
stands for the charge density. The energy-momentum tensor in this
theory is non-conserved as a result of interaction between matter
source and geometrical terms. It causes to generate an extra force
due to which the massive particles are forced to stop their geodesic
motion in the gravitational field. The equation corresponding to the
non-zero divergence of fluid configuration is given by
\begin{eqnarray}\nonumber
\nabla^{\phi}T_{\phi\sigma}&=&\frac{f_{T}(G,T)}{8\pi-f_{T}(G,T)}
\bigg[-\frac{1}{2}g_{\phi\sigma}\nabla^{\phi}T+(\Theta_{\phi\sigma}+T_{\phi\sigma})\nabla^{\phi}(\ln
f_{T}(G,T))\\\label{11}&+&
\nabla^{\phi}\Theta_{\phi\sigma}\bigg].
\end{eqnarray}

The geometrical structure under discussion is chosen to be a
spherical symmetric distribution which is separated by a
hypersurface ($\Sigma$) into two regions (i.e., outer and inner).
The internal composition is discussed by using the metric as
\begin{equation}\label{6}
ds^{2}=-e^{\eta}dt^{2}+e^{\delta}dr^{2}+r^{2}d\theta^{2}+r^2{\sin^{2}\theta}{d\phi^2},
\end{equation}
where $\eta=\eta(r)$ and $\delta=\delta(r)$. The contravariant
component of four-velocity and four-vector in terms of metric
potentials are delineated as
\begin{equation}\label{6a}
\textsf{W}^{\phi}=\left(e^{\frac{-\eta}{2}},0,0,0\right),\quad\textsf{X}^{\phi}
=\left(0,e^{\frac{-\delta}{2}},0,0\right).
\end{equation}
The feasibility and stability of anisotropic compact entities are
observed by utilizing a minimal $f(G,T)$ model \cite{19,21},
prescribed as
\begin{equation}\label{60}
f(G,T)= \mathfrak{t_1}(G)+\mathfrak{t_2}(T),
\end{equation}
where it can be easily seen that $\mathfrak{t_1}$ and
$\mathfrak{t_2}$ depend only on $G$ and $T$, respectively. The vital
role of matter and geometry coupling in comprehending the
substantial features of astrophysical compact structures is devised
through a quadratic model. For this reason,
$\mathfrak{t_1}(G)=\gamma G^2$ with $\gamma$ being the real number
and for the sake of simplicity we will choose its value to be 1
during the graphical analysis (which will be done later) of stars
and $\mathfrak{t_2}(T)=\varrho T$.

Equations \eqref{4} and \eqref{5} are used to write down the field
equations in view of the metric coefficients as
\begin{align}\nonumber
\mathbb{U}&=\frac{1}{48 \pi  r^4}\{e^{-2 \delta} \big(48 r^2 G''-48
r^2 e^{\delta} G''+6 r^2 \delta ' \big(4 \big(e^{\delta}-3\big) G'+r
e^{\delta}\big)+3 r^4 G^2 e^{2 \delta}\\\nonumber&+12 r^2 G \big(2
\big(e^{\delta}-1\big) \eta ''-\big(e^{\delta}-3\big) \eta '\delta
'+\big(e^{\delta}-1\big) \eta '^2\big)+\varrho \mathbb{P}_{r} r^4
e^{2 \delta}+2 \varrho \mathbb{P}_{t} r^4 e^{2 \delta}\\\label{8}&-6
q^2 e^{2 \delta}-9 \varrho \mathbb{U}  r^4 e^{2 \delta}-6 r^2
e^{\delta}+6 r^2 e^{2 \delta}\big)\}^{-1},\\\nonumber
\mathbb{P}_{r}&=\frac{1}{48 \pi r^4}\{e^{-2 \delta} \big(6 r^2 \eta
'\big(4 \big(e^{\delta}-3\big) G'+r e^{\delta}\big)-3 r^4 G^2 e^{2
\delta}-12 r^2 G \big(2 \big(e^{\delta}-1\big) \eta ''
\\\nonumber&-\big(e^{\delta}-3\big) \eta ' \delta '+\big(e^{\delta}-1\big) \eta
'^2\big)+e^{\delta} \big(e^{\delta} \big(-7 \varrho \mathbb{P}_{r}
r^4-2 \varrho \mathbb{P}_{t} r^4+6 q^2+3 \varrho \mathbb{U}
r^4\\\label{9}&-6 r^2\big)+6 r^2\big)\big)\}^{-1},\\\nonumber
\mathbb{P}_{t}&=\frac{1}{96 \pi r^4}\{e^{-2 \delta} \big(-2 \big(-3
r^3 \eta '' \big(r e^{\delta}-8 G'\big)+e^{2 \delta} \big(\varrho
r^4 (\mathbb{P}_{r}+8 \mathbb{P}_{t}-3 \mathbb{U} )+6
q^2\big)\\\nonumber&+3 r^3 e^{\delta} \delta '\big)+3 r^3 \eta '^2
\big(r e^{\delta}-8 G'\big)-3 r^3 \eta ' \big(\delta ' \big(r
e^{\delta}-24 G'\big)-2 \big(e^{\delta}-8
G''\big)\big)\\\label{10}&-6 r^4 G^2 e^{2 \delta}-24 r^2 G \big(2
\big(e^{\delta}-1\big) \eta ''-\big(e^{\delta}-3\big) \eta ' \delta
'+\big(e^{\delta}-1\big) \eta '^2\big)\big)\}^{-1},
\end{align}
where prime indicates the differential with respect to $r$. One can
also notice that the right-hand side of the above equations contains
the matter variables, their involvement is due to the respective
modified theory and hence represents a complex system. The
expressions for the Gauss-Bonnet invariant along with its higher
derivatives take the following form
\begin{align}\label{10a}
G&=\frac{1}{r^2}(2 e^{-2 \delta} \big(-2 \big(e^{\delta}-1\big)
{\eta''}+\big(e^{\delta}-3\big) {\eta'}
{\delta'}-\big(e^{\delta}-1\big) {\eta'}^2\big)),\\\nonumber
G'&=\frac{1}{r^3}\{2 e^{-2 \delta} \big(-2 r \eta ^{'''}
\big(e^{\delta}-1\big)+{\eta''} \big(r \big(3 e^{\delta}-7\big)
{\delta'}+4 \big(e^{\delta}-1\big)\big)\\\nonumber&+{\eta'}^2 \big(r
\big(e^{\delta}-2\big) {\delta'}+2
\big(e^{\delta}-1\big)\big)+{\eta'} \big(r
\big(\big(e^{\delta}-3\big) {\delta''}-2 \big(e^{\delta}-1\big)
{\eta''}\big)\\\label{10b}&+r \big(-\big(e^{\delta}-6\big)\big)
{\delta'}^2-2 \big(e^{\delta}-3\big) {\delta'}\big)\big)\}^{-1},
\\\nonumber G''&=-\frac{1}{r^4}\{2 e^{-2
\delta} \big(2 {\eta''} \big(-r^2 \big(2 e^{\delta}-5\big)
{\delta''}+2 r^2 \big(e^{\delta}-5\big) {\delta'}^2+2 r \big(3
e^{\delta}-7\big) {\delta'}\\\nonumber&+6
\big(e^{\delta}-1\big)\big)+2 r^2 \big(e^{\delta}-1\big)
{\eta''}^2+{\eta'}^2 \big(-r^2 \big(e^{\delta}-2\big) {\delta''}+r^2
\big(e^{\delta}-4\big) {\delta'}^2\\\nonumber&+4 r
\big(e^{\delta}-2\big) {\delta'}+6
\big(e^{\delta}-1\big)\big)-{\eta'} \big({\delta'} \big(4 r^2
\big(e^{\delta}-2\big) {\eta''}-3 \big(r^2 \big(e^{\delta}-6\big)
{\delta''}\\\nonumber&-2 e^{\delta}+6\big)\big)+r^2
\big(e^{\delta}-12\big) {\delta'}^3-r \big(r \big(2 \mu ^{'''}
\big(e^{\delta}-1\big)-\big(e^{\delta}-3\big) \nu
^{'''}\big)\\\nonumber&-8 \big(e^{\delta}-1\big) {\eta''}+4
\big(e^{\delta}-3\big) {\delta''}\big)+4 r \big(e^{\delta}-6\big)
{\delta'}^2\big)-r \big(\mu ^{'''} \big(r \big(5 e^{\delta}-11\big)
{\delta'}\\\label{10c}&+8 \big(e^{\delta}-1\big)\big)-2 r \mu
^{''''} \big(e^{\delta}-1\big)\big)\big)\}^{-1}.
\end{align}
The quantities $\mathbb{U}$, $\mathbb{P}_{r}$ and $\mathbb{P}_{t}$,
after substituting the values of $G,~G'$ and $G''$, turn out to be
\begin{align}\nonumber
\mathbb{U}&=\frac{1}{48 \pi r^6}\{e^{-4 \delta} \big(\varrho
\mathbb{P}_{r} r^6 e^{4 \delta}+2 \varrho  \mathbb{P}_{t} r^6 e^{4
\delta}-6 q^2 r^2 e^{4 \delta}-9 \varrho  \mathbb{U}  r^6 e^{4
\delta}-6 r^4 e^{3 \delta}\\\nonumber&+6 r^4 e^{4 \delta}+6 r \delta
'\big(r \big(r^3 e^{3 \delta}-32 \eta ^{'''} \big(3 e^{2 \delta}-10
e^{\delta}+7\big)\big)+32 \big(7 e^{2 \delta}-24
e^{\delta}+17\big)\\\nonumber&\times \eta ''\big)-384 r^2 \eta
^{''''}e^{\delta}+192 r^2 \eta ^{''''} e^{2 \delta}+192 r^2 \eta
^{''''}+1344 r^2 e^{\delta} \eta '' \delta ''-384 r^2 e^{2 \delta}
\eta '' \delta ''\\\nonumber&-960 r^2 \eta ''\delta ''+48 r^2
\big(11 e^{2 \delta}-64 e^{\delta}+61\big) \eta ''\delta '^2-288 r^2
e^{\delta} \eta ''^2+144 r^2 e^{2 \delta} \eta ''^2\\\nonumber&+144
r^2 \eta ''^2+24 r^2 \big(e^{2 \delta}-4 e^{\delta}+3\big) \eta '^3
\delta '-12 r^2 \big(e^{\delta}-1\big)^2 \eta '^4+12 \eta '^2
\big(-4 \big(e^{\delta}\\\nonumber&-1\big) \big(r^2
\big(e^{\delta}-1\big) \eta ''+2 \big(r^2 \big(e^{\delta}-2\big)
\delta ''-6 e^{\delta}+6\big)\big)+r^2 \big(11 e^{2 \delta}-54
e^{\delta}+47\big) \delta '^2\\\nonumber&+8 r \big(-16 e^{\delta}+5
e^{2 \delta}+11\big) \delta '\big)-48 \eta ' \big(\delta ' \big(r^2
\big(-28 e^{\delta}+9 e^{2 \delta}+19\big) \eta ''-r^2 \big(7 e^{2
\delta}\\\nonumber&-48 e^{\delta}+45\big) \delta ''+12 \big(-4
e^{\delta}+e^{2 \delta}+3\big)\big)+r^2 \big(-35 e^{\delta}+3 e^{2
\delta}+42\big) \delta '^3-2 r \big(e^{\delta}\\\nonumber&-1\big)
\big(r \big(2 \eta ^{'''}
\big(e^{\delta}-1\big)-\big(e^{\delta}-3\big) \delta ^{'''}\big)-8
\big(e^{\delta}-1\big) \eta ''+4 \big(e^{\delta}-3\big) \delta
''\big)+2 r \big(5 e^{2 \delta}\\\nonumber&-34 e^{\delta}+33\big)
\delta '^2\big)+1536 r \eta ^{'''} e^{\delta}-768 r \eta ^{'''}e^{2
\delta}-768 r \eta ^{'''}-2304 e^{\delta} \eta ''\\\label{11a}&+1152
e^{2 \delta} \eta ''+1152 \eta ''\big)\}^{-1},\\\nonumber
\mathbb{P}_{r}&=\frac{1}{48 \pi r^5}\{e^{-4 \delta} \big(r \big(e^{3
\delta} \big(e^{\delta} \big(-7 \varrho \mathbb{P}_{r} r^4-2 \varrho
\mathbb{P}_{t} r^4+6 q^2+3 \varrho \mathbb{U} r^4-6 r^2\big)+6
r^2\big)\\\nonumber&+48 \big(e^{\delta}-1\big)^2 \eta ''^2\big)+6
\eta ' \big(r \big(r^3 e^{3 \delta}-16 \eta ^{'''} \big(e^{2
\delta}-4 e^{\delta}+3\big)\big)+16 \big(e^{\delta}-3\big) \eta
''\\\nonumber&\times\big(r \big(e^{\delta}-3\big) \delta '+2
\big(e^{\delta}-1\big)\big)\big)+24 \big(e^{\delta}-3\big) \eta '^3
\big(r \big(e^{\delta}-3\big) \delta '+4
\big(e^{\delta}-1\big)\big)\\\nonumber&+12 r
\big(e^{\delta}-1\big)^2 \eta '^4-12 \eta '^2 \big(4 r \big(\big(-6
e^{\delta}+e^{2 \delta}+5\big) \eta ''-\big(e^{\delta}-3\big)^2
\delta ''\big)+8 \big(e^{\delta}\\\label{11b}&-3\big)^2 \delta '+3 r
\big(-10 e^{\delta}+e^{2 \delta}+21\big) \delta
'^2\big)\big)\}^{-1},
\\\nonumber
\mathbb{P}_{t}&=\frac{1}{96 \pi  r^5}\{e^{-4 \delta} \big(-2 r
\big(e^{4 \delta} \big(\varrho  r^4 (\mathbb{P}_{r}+8
\mathbb{P}_{t}-3 \mathbb{U} )+6 q^2\big)-3 r \big(r^3 e^{3
\delta}+32 \eta ^{'''}\\\nonumber&\times\big(e^{\delta}-1\big)\big)
\eta ''+3 r \delta ' \big(r^2 e^{3 \delta}+16 \big(3
e^{\delta}-7\big) \eta ''^2\big)-48 \big(e^{2 \delta}-6
e^{\delta}+5\big) \eta ''^2\big)\\\nonumber&-48 \eta '^3 \big(2 r^2
\big(1-e^{\delta}\big) \eta ''+3 r^2 e^{\delta} \delta ''-7 r^2
\delta ''-2 r^2 \big(3 e^{\delta}-10\big) \delta '^2+r \big(e^{2
\delta}\\\nonumber&-20 e^{\delta}+31\big) \delta '-12
e^{\delta}+12\big)+3 \eta '^2 \big(r \big(r \big(r^3 e^{3 \delta}+96
\eta ^{'''} \big(e^{\delta}-1\big)-32
\big(e^{\delta}\\\nonumber&-3\big) \delta ^{'''}\big)+32 \big(-14
e^{\delta}+e^{2 \delta}+13\big) \eta ''+128 \big(e^{\delta}-3\big)
\delta ''\big)-16 \delta ' \big(r^2 \big(19
e^{\delta}\\\nonumber&-33\big) \eta ''-9 r^2 \big(e^{\delta}-5\big)
\delta ''+12 \big(e^{\delta}-3\big)\big)-16 r^2 \big(5
e^{\delta}-42\big) \delta '^3+8 r \big(e^{2 \delta}\\\nonumber&-34
e^{\delta}+141\big) \delta '^2\big)-3 \eta ' \big(r \delta '\big(r
\big(r^3 e^{3 \delta}+64 \eta ^{'''}\big(4 e^{\delta}-7\big)\big)+32
\big(e^{2 \delta}-24 e^{\delta}\\\nonumber&+43\big) \eta ''\big)-16
r^2 \big(19 e^{\delta}-73\big) \eta '' \delta '^2-2 \big(-16 \eta ''
\big(r^2 \big(5 e^{\delta}-13\big) \delta ''-12
\big(e^{\delta}\\\nonumber&-1\big)\big)+64 r^2
\big(e^{\delta}-1\big) \eta ''^2+r \big(r \big(r^2 e^{3 \delta}+32
\eta ^{(4)} \big(e^{\delta}-1\big)\big)-128 \eta ^{'''}
\big(e^{\delta}\\\label{11c}&-1\big)\big)\big)\big)-24 r \eta '^4
\big(2 r \big(e^{\delta}-2\big) \delta '+6 e^{\delta}-e^{2
\delta}-5\big)\big)\}^{-1}.
\end{align}

The structure formation of celestial bodies can be observed by
employing the hydrostatic equilibrium equation, whose expression in
$f(G,T)$ theory is given as
\begin{align}\nonumber
&\frac{\eta'}{2}(\mathbb{U}+\mathbb{P}_{r})+\frac{d}{dr}\big(\mathbb{P}_{r}-\frac{q^2}{8\pi
r^4}\big)
+\frac{2}{r}\big(\mathbb{P}_{r}-\mathbb{P}_{t}-\frac{q^2}{8\pi
r^4}\big)\\\label{12}&
-\frac{\varrho}{8\pi-\varrho}[(\mathbb{P}-2\mathbb{P}_{r})'-\frac{1}{2}(-\mathbb{U}+\mathbb{P}_{r}
+2\mathbb{P}_{t})']=0.
\end{align}
The physical attributes of compact structures are better
comprehended by considering several restrictions, like EoS. The
matter variables of the fluid distribution are associated with this
limitation. One of the fascinating entities present in the universe
is neutron stars, whose occurrence is ascertained through the
collapse of gigantic bodies ranging from eight to twenty times more
massive than the Sun. The conversion to quark stars or black holes
mainly depends upon the lesser and higher densities. Regardless of
their moderate size, a strong gravitational field is created as a
result of these dense stars. The field equations
\eqref{11a}-\eqref{11c} are non-linear incorporating six unknowns,
i.e., metric coefficients, state variables and charge term,
indicating that some limitations must be imposed to solve this
system. To do so, we utilize the MIT bag model EoS in which the
matter variables in the interior region of stellar stars are
interconnected. A pivotal role in disclosing the fundamental
features of strange stars is played by this EoS. Thus, we proceed by
characterizing the quark pressure as
\begin{equation}\label{11}
\mathbb{P}_{r}=\sum_{l}\mathbb{P}^{l}-\mathfrak{B}_{c}, \quad
l=u,d,s,
\end{equation}
where $\mathfrak{B}_{c}$ depicts the bag constant.

The quark matter source is distributed into up, down and strange
terms with their respective pressures as $\mathbb{P}^{u}$,
$\mathbb{P}^{d}$ and $\mathbb{P}^{s}$. Moreover, the relation
$\mathbb{U}^{l}=3\mathbb{P}^{l}$ portrays the association of energy
density and pressure for each quark star. The quark density is
defined in a similar way to the quark pressure by
\begin{equation}\label{12}
\mathbb{U}=\sum_{l}\mathbb{U}^{l}+\mathfrak{B}_{c}, \quad l=u,d,s.
\end{equation}
Equations \eqref{11} and \eqref{12} are employed to formulate the
MIT bag model EoS that helps in revealing the strange matter through
\begin{equation}\label{13}
\mathbb{P}_{r}=\frac{1}{3}(\mathbb{U}-4\mathfrak{B}_{c}).
\end{equation}
For several values of the bag constant, this EoS has been exploited
by many researchers to uncover the substantial attributes of strange
stars \cite{25}.

The main purpose of our work is to determine the anisotropic
solution of the field equations using EoS \eqref{13}. We then
explore its viability with the help of four celestial compact stars.
This notion is accomplished by utilizing the metric coefficients of
a well-known solution, i.e., Tolman IV whose expressions are
described as
\begin{equation}\label{14}
e^{\eta}=\mathfrak{C}^{2}(1+\frac{r^2}{\mathfrak{X}^2}), \quad\quad
e^{\delta}=\frac{1+\frac{2r^2}{\mathfrak{X}^2}}{(1+\frac{r^2}
{\mathfrak{X}^2})(1-\frac{r^2}{\mathfrak{Z}^2})},
\end{equation}
where the unknown parameters for this solution are represented by
$\mathfrak{X}^2$, $\mathfrak{C}^2$ and $\mathfrak{Z}^2$. The values
of these unknowns are found through the junction of outer and inner
geometries. The expressions of state determinants in terms of
$\mathfrak{B}_{c}$ are provided in \textbf{Appendix A}.

\section{Junction Conditions}

The development of anisotropic astrophysical entities can be
understood in a better way by a smooth matching of internal and
external geometries. The metric \eqref{6} has been used to describe
the inner framework, while the outer regime is delineated by
employing the Riessner-Nordstr$\ddot{o}$m spacetime given as
\begin{equation}\label{15}
ds^2=-\bigg(1-\frac{2\mathcal{M}}{r}+\frac{\mathcal{Q}^2}{r^2}\bigg)dt^2
+\bigg(1-\frac{2\mathcal{M}}{r}+\frac{\mathcal{Q}^2}{r^2}\bigg)^{-1}dr^2
+r^2d\theta^{2}+r^2{\sin^{2}\theta}{d\phi^2},
\end{equation}
where $\mathcal{M}$ and $\mathcal{Q}$ indicate the mass and charge
of the outward region, respectively, at the boundary
$r=\mathfrak{R}$. The metric functions present in the spacetimes
\eqref{6} and \eqref{15} are continuous at the junction, resulting
in the following forms
\begin{align}\label{16}
g_{tt}&=\mathfrak{C}^{2}\bigg(1+\frac{\mathfrak{R}^2}{\mathfrak{X}^2}\bigg)
=1-\frac{2\mathcal{M}}{\mathfrak{R}}+\frac{\mathcal{Q}^2}{\mathfrak{R}^2},\\\label{17}
g_{rr}&=\frac{(1+\frac{\mathfrak{R}^2}{\mathfrak{X}^2})
(1-\frac{\mathfrak{R}^2}{\mathfrak{Z}^2})}
{1+\frac{2\mathfrak{R}^2}{\mathfrak{X}^2}}
=1-\frac{2\mathcal{M}}{\mathfrak{R}}+\frac{\mathcal{Q}^2}
{\mathfrak{R}^2},\\\label{18} \frac{\partial g_{tt}}{\partial
r}&=\eta'=\frac{2\mathfrak{R}}{\mathfrak{X}^2+\mathfrak{R}^2}
=\frac{2(\mathcal{M}\mathfrak{R}-\mathcal{Q}^2)}{\mathfrak{R}(\mathfrak{R}^2-2\mathcal{M}\mathfrak{R}+\mathcal{Q}^2)}.
\end{align}
The above equations are solved simultaneously to acquire the values
of $\mathfrak{X}^2$, $\mathfrak{C}^2$ and $\mathfrak{Z}^2$ involved
in Tolman IV solution, which come out to be
\begin{align}\label{19}
\mathfrak{X}^2&=\frac{\mathfrak{R}^3-3 \mathcal{M}
\mathfrak{R}^2}{\mathcal{M}},\\\label{19a}
\mathfrak{C}^2&=\frac{\mathfrak{R}^4 (2 \mathcal{M}-\mathfrak{R})}{2
\mathcal{M}^2 \mathfrak{R}-\mathcal{M} \mathcal{Q}^2-\mathcal{M}
\mathfrak{R}^2+\mathcal{Q}^2 \mathfrak{R}},\\\label{19b}
\mathfrak{Z}^2&=\frac{\mathfrak{R}^4 (2 \mathcal{M}-\mathfrak{R})}{2
\mathcal{M}^2 \mathfrak{R}-\mathcal{M} \mathcal{Q}^2-\mathcal{M}
\mathfrak{R}^2+\mathcal{Q}^2 \mathfrak{R}}.
\end{align}

It is important to mention here that the radial pressure of the
fluid matter distribution must approach to zero on moving towards
the boundary. We utilize the expression of $\mathbb{P}_{r}$ given in
\textbf{Appendix A} to evaluate the radial pressure for the present
modified work as
\begin{align}\nonumber
\mathbb{P}_{r}\mid_{r=\mathfrak{R}}&=\frac{1}{2 (\varrho +8 \pi )
\mathfrak{R}^{16} (\mathcal{M}-\mathfrak{R})^3 (\mathfrak{R}-2
\mathcal{M})^2}\{-4 \mathcal{M}^5 \big(\mathfrak{R}^8 \big(2
\mathfrak{B}_{c} (\varrho +8 \pi )
\\\nonumber&\times\mathfrak{R}^8+21 \mathfrak{R}^6+76992\big)+47808
\mathcal{Q}^8+511872 \mathcal{Q}^6 \mathfrak{R}^2+1057920
\mathcal{Q}^4 \mathfrak{R}^4\\\nonumber&+\mathcal{Q}^2
\mathfrak{R}^6 \big(3 \mathfrak{R}^6+602560\big)\big)+2
\mathcal{M}^4 \big(\mathfrak{R}^9 \big(16 \mathfrak{B}_{c} (\varrho
+8 \pi ) \mathfrak{R}^8+57
\mathfrak{R}^6\\\nonumber&+9984\big)+107136 \mathcal{Q}^8
\mathfrak{R}+514560 \mathcal{Q}^6 \mathfrak{R}^3+589504
\mathcal{Q}^4 \mathfrak{R}^5+20 \mathcal{Q}^2
\big(\mathfrak{R}^6\\\nonumber&+9184\big)
\mathfrak{R}^7\big)-\mathcal{M}^3 \big(25 \mathfrak{R}^{16} \big(2
\mathfrak{B}_{c} (\varrho +8 \pi ) \mathfrak{R}^2+3\big)+88576
\mathcal{Q}^8 \mathfrak{R}^2\\\nonumber&+237568 \mathcal{Q}^6
\mathfrak{R}^4+151232 \mathcal{Q}^4 \mathfrak{R}^6+\mathcal{Q}^2
\big(51 \mathfrak{R}^6+20288\big) \mathfrak{R}^8\big)+\mathcal{M}^2
\\\nonumber&\times\big(2 \mathfrak{R}^{17} \big(19 \mathfrak{B}_{c} (\varrho +8 \pi )
\mathfrak{R}^2+12\big)+15872 \mathcal{Q}^8 \mathfrak{R}^3+24064
\mathcal{Q}^6 \mathfrak{R}^5\\\nonumber&+6848 \mathcal{Q}^4
\mathfrak{R}^7+31 \mathcal{Q}^2 \mathfrak{R}^{15}\big)-\mathcal{M}
\big(\mathfrak{R}^{18} \big(14 \mathfrak{B}_{c} (\varrho +8 \pi )
\mathfrak{R}^2+3\big)\\\nonumber&+768 \mathcal{Q}^6
\mathfrak{R}^6+1024 \mathcal{Q}^8 \mathfrak{R}^4+9 \mathcal{Q}^2
\mathfrak{R}^{16}\big)+\mathfrak{R}^{17} \big(2 \mathfrak{B}_{c}
(\varrho +8 \pi )
\mathfrak{R}^4+\mathcal{Q}^2\big)\\\nonumber&-3059712 \mathcal{M}^9
\mathfrak{R}^4+24576 \mathcal{M}^8 \big(249 \mathcal{Q}^2
\mathfrak{R}^3+248 \mathfrak{R}^5\big)-6144 \mathcal{M}^7
\\\nonumber&\times\big(747 \mathcal{Q}^4 \mathfrak{R}^2+1767 \mathcal{Q}^2
\mathfrak{R}^4+767 \mathfrak{R}^6\big)+8 \mathcal{M}^6 \big(191232
\mathcal{Q}^6 \mathfrak{R}\\\label{22}&+892800 \mathcal{Q}^4
\mathfrak{R}^3+926464 \mathcal{Q}^2 \mathfrak{R}^5+3
\big(\mathfrak{R}^6+72960\big) \mathfrak{R}^7\big)\}^{-1}=0.
\end{align}
With the help of Eq.\eqref{22}, we have the expression of
$\mathfrak{B}_{c}$ represented as
\begin{align}\nonumber
\mathfrak{B}_{c}&=\frac{1}{2 (\varrho +8 \pi ) \mathfrak{R}^{16}
(\mathcal{M}-\mathfrak{R})^3 (\mathfrak{R}-2 \mathcal{M})^2}\{24576
\mathcal{M}^8 \big(249 \mathcal{Q}^2 \mathfrak{R}^3\\\nonumber&+248
\mathfrak{R}^5\big)-3059712 \mathcal{M}^9 \mathfrak{R}^4-6144
\mathcal{M}^7 \big(747 \mathcal{Q}^4 \mathfrak{R}^2+1767
\mathcal{Q}^2 \mathfrak{R}^4\\\nonumber&+767 \mathfrak{R}^6\big)+8
\mathcal{M}^6 \big(191232 \mathcal{Q}^6 \mathfrak{R}+892800
\mathcal{Q}^4 \mathfrak{R}^3+926464 \mathcal{Q}^2
\mathfrak{R}^5\\\nonumber&+3 \big(\mathfrak{R}^6+72960\big)
\mathfrak{R}^7\big)-4 \mathcal{M}^5 \big(47808 \mathcal{Q}^8+511872
\mathcal{Q}^6 \mathfrak{R}^2+1057920\\\nonumber&\times \mathcal{Q}^4
\mathfrak{R}^4+\mathcal{Q}^2 \mathfrak{R}^6 \big(3
\mathfrak{R}^6+602560\big)+3 \mathfrak{R}^8 \big(7
\mathfrak{R}^6+25664\big)\big)+2 \mathcal{M}^4\\\nonumber&\times
\big(107136 \mathcal{Q}^8 \mathfrak{R}+514560 \mathcal{Q}^6
\mathfrak{R}^3+589504 \mathcal{Q}^4 \mathfrak{R}^5+20 \mathcal{Q}^2
\big(\mathfrak{R}^6\\\nonumber&+9184\big) \mathfrak{R}^7+3 \big(19
\mathfrak{R}^6+3328\big) \mathfrak{R}^9\big)-\mathcal{M}^3
\big(88576 \mathcal{Q}^8 \mathfrak{R}^2+237568 \mathcal{Q}^6
\mathfrak{R}^4\\\nonumber&+151232 \mathcal{Q}^4
\mathfrak{R}^6+\mathcal{Q}^2 \big(51 \mathfrak{R}^6+20288\big)
\mathfrak{R}^8+75 \mathfrak{R}^{16}\big)+\mathcal{M}^2 \big(15872
\mathcal{Q}^8 \mathfrak{R}^3\\\nonumber&+24064 \mathcal{Q}^6
\mathfrak{R}^5+6848 \mathcal{Q}^4 \mathfrak{R}^7+31 \mathcal{Q}^2
\mathfrak{R}^{15}+24 \mathfrak{R}^{17}\big)-\mathcal{M} \big(1024
\mathcal{Q}^8 \mathfrak{R}^4\\\label{23}&+768 \mathcal{Q}^6
\mathfrak{R}^6+9 \mathcal{Q}^2 \mathfrak{R}^{16}+3
\mathfrak{R}^{18}\big)+\mathcal{Q}^2 \mathfrak{R}^{17}\}^{-1}.
\end{align}
The values of $\mathfrak{X}^2$, $\mathfrak{C}^2$, $\mathfrak{Z}^2$
and $\mathfrak{B}_{c}$ are determined using the experimental data
from four distinct stellar candidates, including their radii and
masses (Table \textbf{1}). Interestingly, one can notice the
consistent behavior of these compact entities with the Buchdhal
limit \cite{22c}, i.e.,
\begin{equation*}
\frac{2\mathcal{M}}{\mathfrak{R}}<\frac{8}{9}.
\end{equation*}
\begin{table}
\scriptsize \centering \caption{Physical values (masses and radii)
of four stellar bodies} \label{Table1} \vspace{+0.1in}
\setlength{\tabcolsep}{0.95em}
\begin{tabular}{cccccc}
\hline\hline \\Stellar Objects & Her X-I & LMC X-4 & 4U1820-30 &PSR
J1614-2230
\\\hline \\
Mass $(\mathcal{M}_{\odot})$ & 0.85 & 1.04 & 1.58 & 1.97
\\\hline \\
$\mathfrak{R}$ (km) & 8.1 & 8.3 & 9.1 & 9.69
\\\hline \\
$\mathcal{M}/\mathfrak{R}$ & 0.1546 & 0.1846 & 0.2559& 0.2996 \\
\hline\hline
\end{tabular}
\end{table}
The bag constant value is derived from Eq.\eqref{23} in order to
examine the stellar progression effectively. The masses and radii of
the models under consideration are used to ascertain the values of
$\mathfrak{B}_c$ and unknowns of Tolman IV solution and their values
are specified in Table \textbf{2} and \textbf{3}, respectively, for
both charges.
\begin{table}
\scriptsize \centering \caption{Approximated values of constants
$\mathfrak{X}^2$, $\mathfrak{C}^2$, $\mathfrak{Z}^2$ and
$\mathfrak{B}_{c}$ values for $\mathcal{Q}=0.1$} \label{Table1}
\vspace{+0.1in} \setlength{\tabcolsep}{0.95em}
\begin{tabular}{cccccc}
\hline\hline \\Stellar Objects & Her X-I & LMC X-4 & 4U1820-30 &PSR
J1614-2230
\\\hline \\
$\mathfrak{X}^2$ & 227.339 & 166.325 & 75.141 & 28.2042
\\\hline \\
$\mathfrak{C}^2$ & 0.536081 & 0.446021 & 0.232282 & 0.0925942
\\\hline \\
$\mathfrak{Z}^2$ & 424.681& 373.375 & 323.804& 304.877
\\\hline \\
$\mathfrak{B}_{c}$ & 0.000112688 & 0.000121328 & 0.000118646 & 0.000108779\\
\hline\hline
\end{tabular}
\end{table}
\begin{table}[H]
\scriptsize \centering \caption{Approximated values of constants
$\mathfrak{X}^2$, $\mathfrak{C}^2$, $\mathfrak{Z}^2$ and
$\mathfrak{B}_{c}$ values for $\mathcal{Q}=1.5$} \label{Table1}
\vspace{+0.1in} \setlength{\tabcolsep}{0.95em}
\begin{tabular}{cccccc}
\hline\hline \\Stellar Objects & Her X-I & LMC X-4 & 4U1820-30 &PSR
J1614-2230
\\\hline \\
$\mathfrak{X}^2$ & 227.339 & 166.325 & 75.141 & 28.2042
\\\hline \\
$\mathfrak{C}^2$ & 0.562576 & 0.469013 & 0.24515 & 0.0982896
\\\hline \\
$\mathfrak{Z}^2$ & 528.14 & 483.549 & 386.043& 355.323
\\\hline \\
$\mathfrak{B}_{c}$ & 0.000106252 & 0.000116292 & 0.000116657 & 0.000108096\\
\hline\hline
\end{tabular}
\end{table}

The range of bag constant is considered to be 58.9-91.5 $MeV/fm^3$
for the quark source having negligible mass \cite{26}, while the
quark candidates possessing mass comparable to 154 $MeV$ have the
range 56-78 $MeV/fm^3$ \cite{27}. It is worth noting that there has
been a large body of literature available in which the larger values
of $\mathfrak{B}_c$ are employed. According to Xu \cite{28}, for
fixed $\mathfrak{B}_{c}=60,~100~MeV/fm^3$, the celestial structure
LMXB EXO 0748-676 can be regarded as quark star. The bag model,
which depends on density, may provide a wide range of values for
this constant, as per several experimental findings reported by
CERN-SPS and RHIC \cite{29}. The values of bag constant derived in
this theory corresponding to the strange quark candidates with
$\mathcal{Q}=0.1$ are 85.15, 91.68, 89.65 and 82.20 $MeV/fm^3$,
respectively and for $\mathcal{Q}=1.5$, their values reduce to
80.28, 87.87, 88.15 and 81.68 $MeV/fm^3$, respectively. It is
remarkable to note that under this scenario, the calculated values
are closely related to the predicted ranges.

\section{Graphical Analysis of Compact Bodies}

Here, we investigate the influence of electromagnetic field and
$f(G,T)$ gravity on anisotropic astrophysical entities by discussing
some of their substantial characteristics. The physical features are
studied by using masses and radii of the assumed strange candidates.
The graphical analysis is done thoroughly with the help of model
\eqref{60} by keeping $\varrho$ and $\gamma$ to be 0.5 and 1,
respectively. Further, all the compact stars are shown graphically
by choosing two particular values of charge, i.e., $\mathcal{Q}=0.1$
and 1.5. For this reason, the smooth lines manifest the behavior of
considered stars with a lower charge value, while the dash lines
stand for the higher charge value in all plots. The corresponding
charge to mass ratios related to four considered stars have been
provided in Table \textbf{4}. This ratio is generally expected to be
very small for astrophysical objects \cite{29a}.
\begin{table}
\scriptsize \centering \caption{Approximated values of
$\frac{\mathcal{Q}}{\mathcal{M}}$ corresponding to four stars}
\label{Table1} \vspace{+0.1in} \setlength{\tabcolsep}{0.95em}
\begin{tabular}{cccccc}
\hline\hline \\Stellar Objects & $\mathcal{Q}=0.1$ &
$\mathcal{Q}=1.5$ &
\\\hline \\
Her X-I & 0.771$\times$10$^{-20}$ & 0.116$\times$10$^{-18}$
\\\hline \\
LMC X-4 & 0.630$\times$10$^{-20}$ & 0.094$\times$10$^{-18}$
\\\hline \\
4U1820-30 & 0.415$\times$10$^{-20}$ & 0.062$\times$10$^{-18}$
\\\hline \\
PSR J1614-2230 & 0.332$\times$10$^{-20}$ & 0.050$\times$10$^{-18}$ \\
\hline\hline
\end{tabular}
\end{table}

First of all, we inspect the feasibility of radial and temporal
metric coefficients, behavior of state variables, anisotropy and
energy constraints. Later, the stability of all the observed quark
candidates is examined. It should be kept in mind that the solution
shows compatible behavior when its metric potentials rise
monotonically with respect to radial coordinate. The metric
potentials given in Eq.\eqref{14} contain three unknown constants,
their values associated with both charges are presented in Tables
\textbf{2} and \textbf{3}. The plot in Figure \textbf{1} confirms
the consistency of the solution for both charge terms. It is
important to mention here that all the plots of compact candidates
Her X-I, LMC X-4, 4U1820-30 and PSR J1614-2230 are illustrated by
green, pink, black and red colors in the case of lower and higher
charge values.
\begin{figure}\center
\epsfig{file=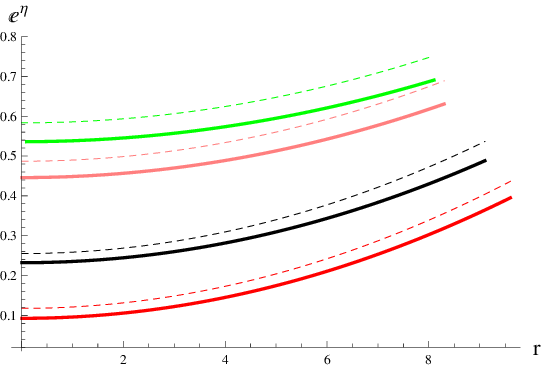,width=0.45\linewidth}\epsfig{file=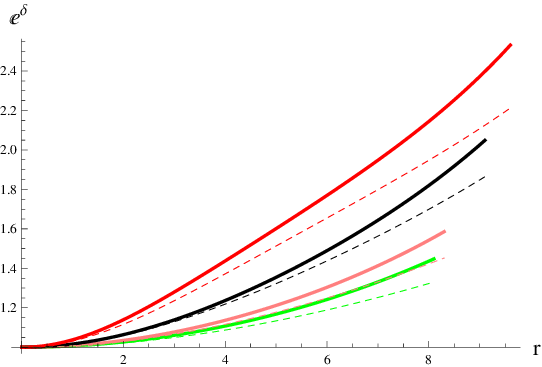,width=0.45\linewidth}
\caption{Plots of metric potentials of Tolman IV versus $r$ for Her
X-I, LMC X-4, 4U1820-30 and PSR J1614-2230 quark candidates.}
\end{figure}

\subsection{Behavior of State Variables}
\begin{figure}\center
\epsfig{file=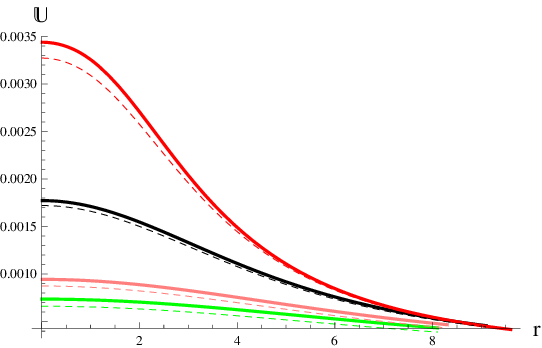,width=0.45\linewidth}\epsfig{file=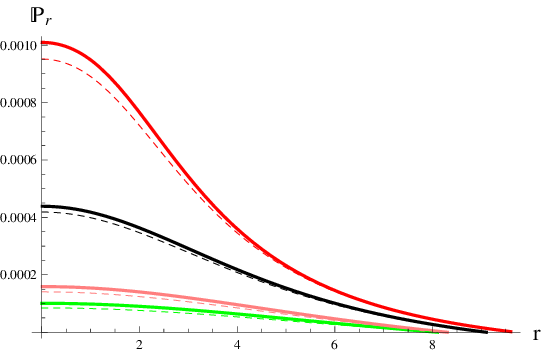,width=0.45\linewidth}
\epsfig{file=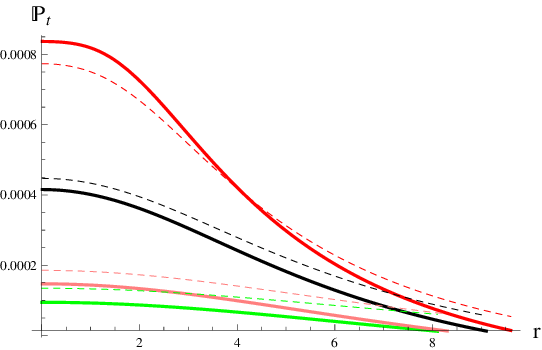,width=0.45\linewidth}
\caption{Analysis of energy density, radial/tangential pressures
versus $r$ for Her X-I, LMC X-4, 4U1820-30 and PSR J1614-2230 quark
candidates.}
\end{figure}

The mandatory role of physical factors (pressure ingredients, energy
density and charge) in immensely dense compact entities cannot be
ignored. In order to determine the physical acceptance of these
celestial objects, we figure out the contribution of physical
variables. The behavior of these quantities guarantees the
acceptability if they remain positive as a whole, display maximal
value at the center and decrease smoothly as radial coordinate
increases. The density profile presented in Figure \textbf{2}
reveals the denser core and deteriorates when $r$ increases. It can
also be noticed that for $\mathcal{Q}=1.5$, the quark stars develop
into less dense composition. The pressure ingredients
$\mathbb{P}_{r}$ and $\mathbb{P}_{t}$ demonstrate the analogous
behavior of density and approach zero at the boundary. The
regularity of the proposed solution is achieved by satisfying some
requirements at the center, which are shown as
\begin{align*}
\frac{d\mathbb{U}}{dr}|_{r=0}=0,\quad
\frac{d\mathbb{P}_{r}}{dr}|_{r=0}=0,\quad\frac{d\mathbb{P}_{t}}{dr}|_{r=0}=0,\\\nonumber
\frac{d^2\mathbb{U}}{dr^2}|_{r=0}<0,\quad
\frac{d^2\mathbb{P}_{r}}{dr^2}|_{r=0}<0,\quad\frac{d^2\mathbb{P}_{t}}{dr^2}|_{r=0}<0.
\end{align*}
From the Figures \textbf{3} and \textbf{4}, one can confirm the
regularity of solution as all the considered strange candidates
fulfill the regularity conditions under the impact of both charges.
\begin{figure}\center
\epsfig{file=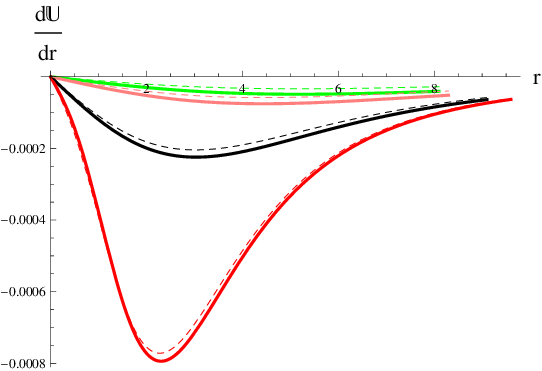,width=0.45\linewidth}\epsfig{file=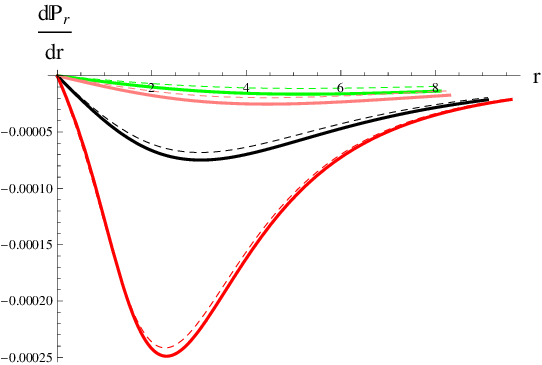,width=0.45\linewidth}
\epsfig{file=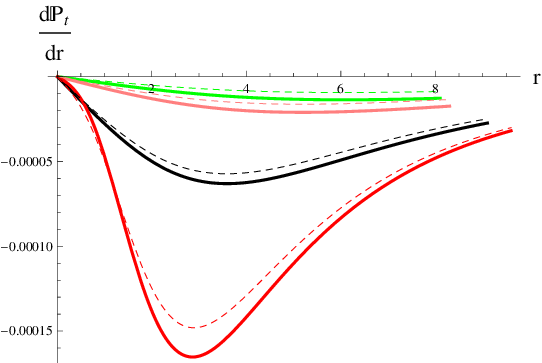,width=0.45\linewidth} \caption{Behavior of
$\frac{d\mathbb{U}}{dr}$, $\frac{d\mathbb{P}_{r}}{dr}$ and
$\frac{d\mathbb{P}_{t}}{dr}$ versus $r$ for Her X-I, LMC X-4,
4U1820-30 and PSR J1614-2230 quark candidates.}
\end{figure}
\begin{figure}\center
\epsfig{file=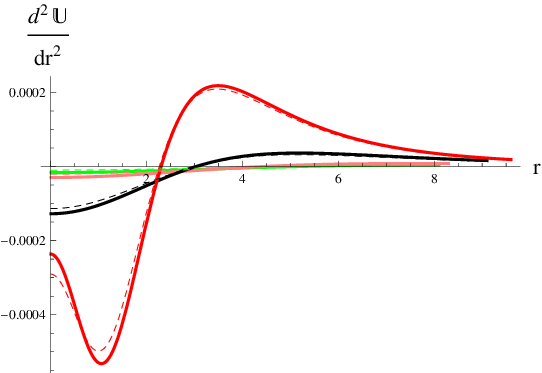,width=0.45\linewidth}\epsfig{file=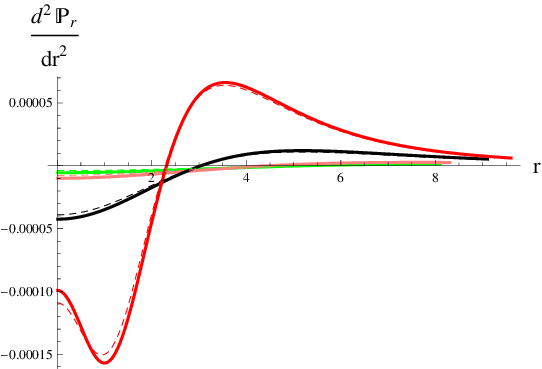,width=0.45\linewidth}
\epsfig{file=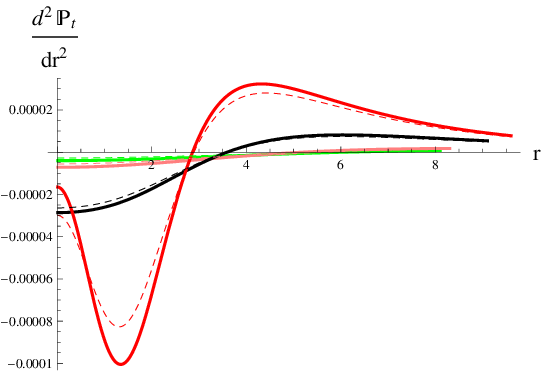,width=0.45\linewidth} \caption{Behavior of
$\frac{d^2\mathbb{U}}{dr^2}$, $\frac{d^2\mathbb{P}_{r}}{dr^2}$ and
$\frac{d^2\mathbb{P}_{t}}{dr^2}$ versus $r$ for Her X-I, LMC X-4,
4U1820-30 and PSR J1614-2230 quark candidates.}
\end{figure}

\subsection{Anisotropic and Essential Factors}

The anisotropy in the system is generated due to the existence of
radial and transversal pressures and is denoted as
$\Delta=\mathbb{P}_{t}-\mathbb{P}_{r}$. The anisotropic factor is
derived on the basis of pressure components (given in
\textbf{Appendix A}) corresponding to Tolman IV solution and bag
constant, leading to following expression
\begin{align}\nonumber
\Delta&=\frac{1}{4 r^4 \big(2 r^2+\mathfrak{X}^2\big)^7
\mathfrak{Z}^8 (\varrho +6 \pi ) (\varrho +8 \pi )}\{3 \big(64
\mathfrak{B}_{c} \pi ^2 r^4 \big(2 r^2+\mathfrak{X}^2\big)^7
\mathfrak{Z}^8\\\nonumber&+\varrho  \big(64 \big(4 \mathfrak{B}_{c}
\varrho \mathfrak{Z}^8-3 \mathfrak{Z}^6+36\big) r^{18}+64 \big(2
\big(7 \mathfrak{B}_{c} \varrho \mathfrak{Z}^8-5
\mathfrak{Z}^6+57\big) \mathfrak{X}^2+\mathfrak{Z}^2
\big(\mathfrak{Z}^6\\\nonumber&-24\big)\big) r^{16}+16 \big(\big(84
\mathfrak{B}_{c} \varrho \mathfrak{Z}^8-57 \mathfrak{Z}^6+528\big)
\mathfrak{X}^4+4 \mathfrak{Z}^2 \big(3 \mathfrak{Z}^6-136\big)
\mathfrak{X}^2-8 \mathfrak{Z}^4 \\\nonumber&\times\big(\mathcal{Q}^2
\mathfrak{Z}^4+22\big)\big) r^{14}+16 \big(\big(70 \mathfrak{B}_{c}
\varrho \mathfrak{Z}^8-45 \mathfrak{Z}^6+592\big) \mathfrak{X}^6+5
\mathfrak{Z}^2 \big(3 \mathfrak{Z}^6\\\nonumber&+104\big)
\mathfrak{X}^4-28 \mathfrak{Z}^4 \big(\mathcal{Q}^2
\mathfrak{Z}^4-46\big) \mathfrak{X}^2+192 \mathfrak{Z}^6\big)
r^{12}+4 \big(\big(140 \mathfrak{B}_{c} \varrho \mathfrak{Z}^8-85
\mathfrak{Z}^6\\\nonumber&+1408\big) \mathfrak{X}^8+8 \mathfrak{Z}^2
\big(5 \mathfrak{Z}^6-72\big) \mathfrak{X}^6-8 \mathfrak{Z}^4
\big(21 \mathcal{Q}^2 \mathfrak{Z}^4+982\big) \mathfrak{X}^4-12224
\mathfrak{Z}^6 \mathfrak{X}^2\big) r^{10}\\\nonumber&+4
\mathfrak{X}^2 \big(\big(42 \mathfrak{B}_{c} \varrho
\mathfrak{Z}^8-24 \mathfrak{Z}^6-512\big)
\mathfrak{X}^8+\mathfrak{Z}^2 \big(15 \mathfrak{Z}^6-5984\big)
\mathfrak{X}^6-28 \mathfrak{Z}^4 \big(5 \mathcal{Q}^2
\mathfrak{Z}^4\\\nonumber&+442\big) \mathfrak{X}^4-3904
\mathfrak{Z}^6 \mathfrak{X}^2+8064 \mathfrak{Z}^8\big)
r^8+\mathfrak{X}^4 \mathfrak{Z}^2 \big(\mathfrak{Z}^4 \big(28
\mathfrak{B}_{c} \mathfrak{Z}^2 \varrho -15\big)
\mathfrak{X}^8\\\nonumber&+4 \big(3 \mathfrak{Z}^6+704\big)
\mathfrak{X}^6-280 \mathfrak{Z}^2 \big(\mathcal{Q}^2
\mathfrak{Z}^4-80\big) \mathfrak{X}^4+47936 \mathfrak{Z}^4
\mathfrak{X}^2+32832 \mathfrak{Z}^6\big)
r^6\\\nonumber&+\mathfrak{X}^6 \mathfrak{Z}^4 \big(\mathfrak{Z}^2
\big(2 \mathfrak{B}_{c} \mathfrak{Z}^2 \varrho -1\big)
\mathfrak{X}^8+\mathfrak{Z}^4 \mathfrak{X}^6-4 \big(21 \mathcal{Q}^2
\mathfrak{Z}^4+184\big) \mathfrak{X}^4-3776 \mathfrak{Z}^2
\mathfrak{X}^2\\\nonumber&-4320 \mathfrak{Z}^4\big) r^4-14
\mathcal{Q}^2 \mathfrak{X}^{12} \mathfrak{Z}^8 r^2-\mathcal{Q}^2
\mathfrak{X}^{14} \mathfrak{Z}^8\big)+4 \pi \big(384 \big(2
\mathfrak{B}_{c} \varrho
\mathfrak{Z}^8-\mathfrak{Z}^6\\\nonumber&+12\big) r^{18}+32
\big(\big(84 \mathfrak{B}_{c} \varrho \mathfrak{Z}^8-41
\mathfrak{Z}^6+480\big) \mathfrak{X}^2+2 \mathfrak{Z}^2
\big(\mathfrak{Z}^6-24\big)\big) r^{16}+16\\\nonumber&\times
\big(\big(252 \mathfrak{B}_{c} \varrho \mathfrak{Z}^8-121
\mathfrak{Z}^6+1088\big) \mathfrak{X}^4+2 \mathfrak{Z}^2 \big(5
\mathfrak{Z}^6-536\big) \mathfrak{X}^2-16 \mathfrak{Z}^4
\big(\mathcal{Q}^2 \mathfrak{Z}^4\\\nonumber&+28\big)\big) r^{14}+32
\big(\big(105 \mathfrak{B}_{c} \varrho \mathfrak{Z}^8-50
\mathfrak{Z}^6+664\big) \mathfrak{X}^6+\mathfrak{Z}^2 \big(5
\mathfrak{Z}^6+848\big) \mathfrak{X}^4-4
\mathfrak{Z}^4\\\nonumber&\times \big(7 \mathcal{Q}^2
\mathfrak{Z}^4-410\big) \mathfrak{X}^2+160 \mathfrak{Z}^6\big)
r^{12}+16 \big(\big(105 \mathfrak{B}_{c} \varrho \mathfrak{Z}^8-50
\mathfrak{Z}^6+1024\big) \mathfrak{X}^8\\\nonumber&+\mathfrak{Z}^2
\big(5 \mathfrak{Z}^6+1088\big) \mathfrak{X}^6-28 \mathfrak{Z}^4
\big(3 \mathcal{Q}^2 \mathfrak{Z}^4+94\big) \mathfrak{X}^4-6464
\mathfrak{Z}^6 \mathfrak{X}^2\big) r^{10}+2
\mathfrak{X}^2\\\nonumber&\times \big(\big(252 \mathfrak{B}_{c}
\varrho \mathfrak{Z}^8-121 \mathfrak{Z}^6-1280\big) \mathfrak{X}^8+2
\mathfrak{Z}^2 \big(5 \mathfrak{Z}^6-11712\big) \mathfrak{X}^6-16
\mathfrak{Z}^4 \\\nonumber&\times\big(35 \mathcal{Q}^2
\mathfrak{Z}^4+3732\big) \mathfrak{X}^4-35392 \mathfrak{Z}^6
\mathfrak{X}^2+25344 \mathfrak{Z}^8\big) r^8+\mathfrak{X}^4
\big(\big(84 \mathfrak{B}_{c} \varrho \mathfrak{Z}^8\\\nonumber&-41
\mathfrak{Z}^6-1024\big) \mathfrak{X}^8+2 \mathfrak{Z}^2
\big(\mathfrak{Z}^6-1408\big) \mathfrak{X}^6+\big(20992
\mathfrak{Z}^4-560 \mathcal{Q}^2 \mathfrak{Z}^8\big)
\mathfrak{X}^4\\\nonumber&+69824 \mathfrak{Z}^6 \mathfrak{X}^2+57600
\mathfrak{Z}^8\big) r^6+3 \mathfrak{X}^6 \mathfrak{Z}^2
\big(\mathfrak{Z}^4 \big(2 \mathfrak{B}_{c} \mathfrak{Z}^2 \varrho
-1\big) \mathfrak{X}^8+256 \mathfrak{X}^6\\\nonumber&-8
\mathfrak{Z}^2 \big(7 \mathcal{Q}^2 \mathfrak{Z}^4-128\big)
\mathfrak{X}^4+448 \mathfrak{Z}^4 \mathfrak{X}^2-960
\mathfrak{Z}^6\big) r^4-28 \mathcal{Q}^2 \mathfrak{X}^{12}
\mathfrak{Z}^8 r^2\\\label{16a}&-2 \mathcal{Q}^2 \mathfrak{X}^{14}
\mathfrak{Z}^8\big)\big)\}^{-1}.
\end{align}
Using the observational data presented in Table \textbf{1}, we
investigate how anisotropy affects the structural progression of
spherical matter source. When the tangential pressure is smaller
than radial pressure ($\mathbb{P}_{t}<\mathbb{P}_{r}$), it indicates
$\Delta<0$ and pressure is applied in the inward direction. The
expression $\Delta>0$ is attained for
$\mathbb{P}_{t}>\mathbb{P}_{r}$ implying that anisotropic pressure
is in the outward direction. For $\mathcal{Q}=0.1$, the anisotropy
of the stars Her X-I, LMC X-4 and 4U1820-30 is negative at first and
then becomes positive, however, PSR J1614-2230 shows negative
anisotropy in the whole domain. When $\mathcal{Q}=1.5$, the stellar
candidates Her X-I and LMC X-4 have positive anisotropic pressure
throughout, whereas the anisotropy of 4U1820-30 and PSR J1614-2230
transforms from negative to positive (Figure \textbf{5}).
\begin{figure}\center
\epsfig{file=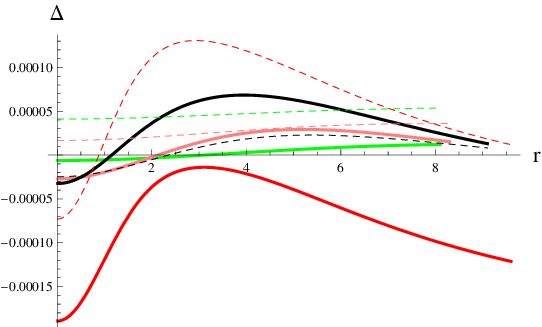,width=0.4\linewidth}\epsfig{file=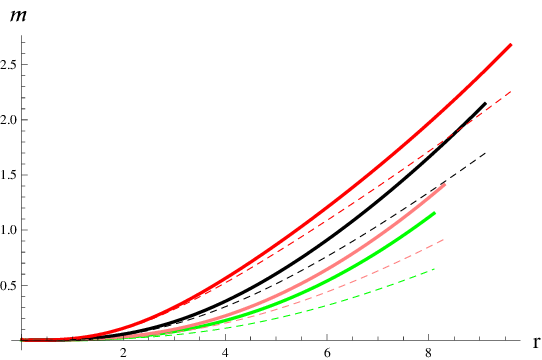,width=0.4\linewidth}
\epsfig{file=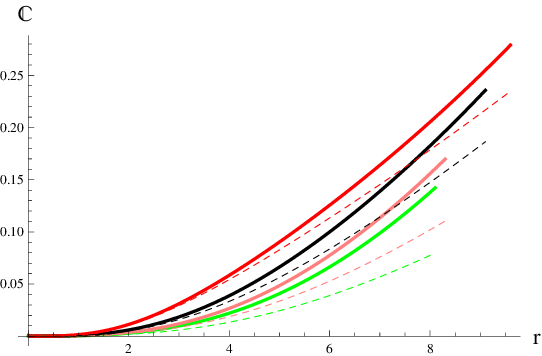,width=0.4\linewidth}\epsfig{file=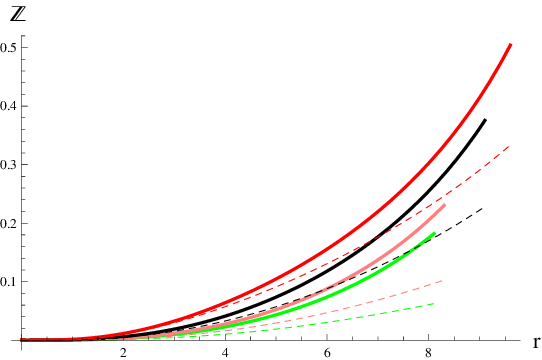,width=0.4\linewidth}
\caption{Plots of anisotropy, mass, compactness and redshift
versus $r$ for Her X-I, LMC X-4, 4U1820-30 and PSR J1614-2230
quark candidates.}
\end{figure}

One of the essential characteristics of spherical symmetric
structure is mass, which is mathematically denoted as
\begin{equation}\label{58}
\mathit{m}=4\pi\int^{\mathfrak{R}}_{0}\mathbb{U}r^2dr,
\end{equation}
where $\mathbb{U}$ is expressed in \textbf{Appendix A}. The mass of
strange stellar bodies is determined through numerical manipulation
of Eq.\eqref{58} with $\mathit{m}(0)=0$ as the initial condition.
Compactness (the ratio of mass and radius) is another important
physical characteristic of celestial objects. Its mathematical
formula is given as $\mathbb{C}=\frac{\mathit{m}}{r}$ and is
graphically portrayed using the previously mentioned condition. In
order to determine the maximum bound of compactness factor, Buchdahl
\cite{22c} matched the outer and inner regimes of spherical system
to conclude that its values should not exceed $\frac{4}{9}$. The
astrophysical objects present in vigorous gravitational pull produce
enlarged electromagnetic radiation, whose increment in wavelength is
measured using redshift factor and is delineated as
\begin{equation*}
\mathbb{Z}=\frac{1}{\sqrt{1-2\mathbb{C}}}-1.
\end{equation*}
On the celestial surface of isotropic matter configuration, the
maximum range of this factor is $\mathbb{Z}<2$, while it becomes
5.211 for the anisotropic configured source. Figure \textbf{5} shows
that these factors (mass, compactness and redshift) comply within
their required bounds for all the considered objects.

\subsection{Energy Constraints}

Several mathematical constraints are used to ensure the occurrence
of usual matter configuration present in the internal regime of
cosmic bodies. These restrictions are known as energy conditions and
have proven to be interesting criterion that differentiate between
exotic and normal matter sources. The accomplishment of these
limitations indicates regular source and signifies the viability of
the resulting solution. These requirements, for the anisotropic
configuration, are classified into null (NEC), dominant (DEC),
strong (SEC) and weak energy condition (WEC), respectively, as
\begin{align}\nonumber
\text{NEC:}\quad&\mathbb{U}+\mathbb{P}_{r}\geq0,\quad\mathbb{U}+\mathbb{P}_{t}+\frac{q^2}{4\pi
r^4}\geq0,
\\\nonumber\text{DEC:}\quad &\mathbb{U}-\mathbb{P}_{t}\geq0,\quad
\mathbb{U}-\mathbb{P}_{r}+\frac{q^2}{4\pi r^4}\geq0,
\\\nonumber\text{SEC:}\quad
&\mathbb{U}+\mathbb{P}_{r}+2\mathbb{P}_{t}+\frac{q^2}{4\pi
r^4}\geq0,
\\\label{59} \text{WEC:}\quad &\mathbb{U}+\frac{q^2}{8\pi
r^4}\geq0, \quad
\mathbb{U}+\mathbb{P}_{r}\geq0,\quad\mathbb{U}+\mathbb{P}_{t}+\frac{q^2}{4\pi
r^4}\geq0.
\end{align}
The energy constraints corresponding to all compact structures under
the influence of electromagnetic field are graphically analyzed in
Figure \textbf{6}. It can be observed that both the solution and
stars depict realistic behavior in the present work for both
$\mathcal{Q}=0.1$ and $1.5$. This shows that all the strange quark
candidates incorporate ordinary matter internally.
\begin{figure}\center
\epsfig{file=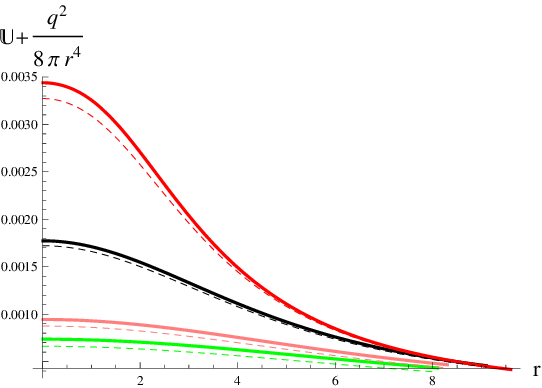,width=0.4\linewidth}\epsfig{file=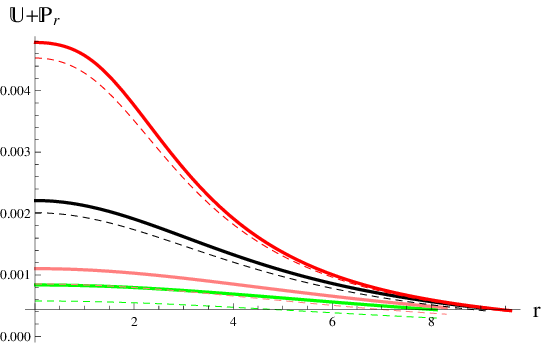,width=0.4\linewidth}
\epsfig{file=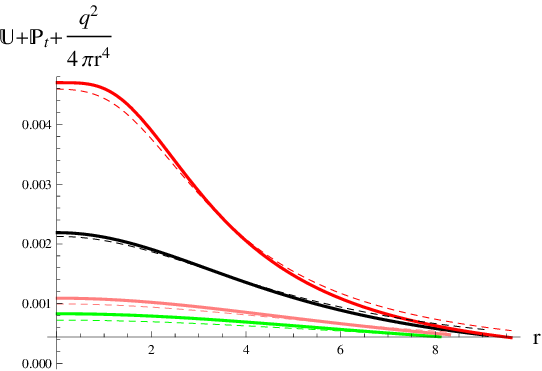,width=0.4\linewidth}\epsfig{file=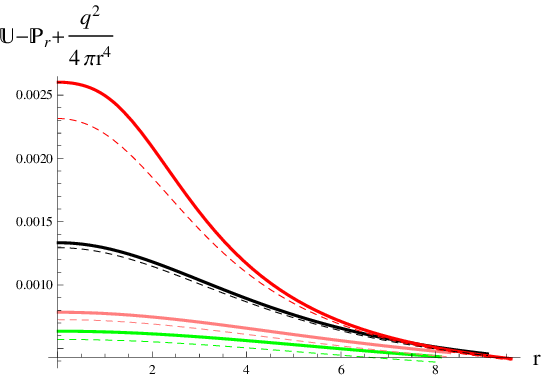,width=0.4\linewidth}
\epsfig{file=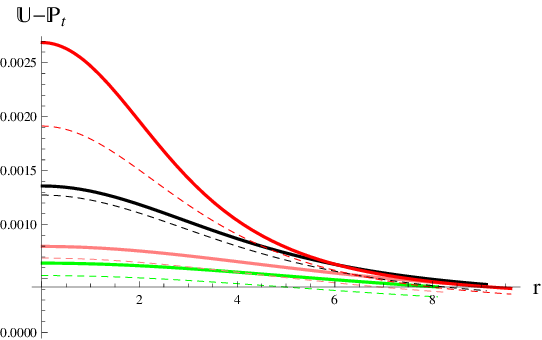,width=0.4\linewidth}\epsfig{file=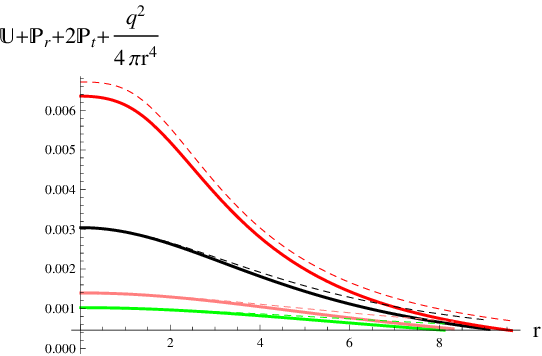,width=0.4\linewidth}
\caption{Behavior of energy constraints versus $r$ for Her X-I, LMC
X-4, 4U1820-30 and PSR J1614-2230 quark candidates.}
\end{figure}

\subsection{Stability Conditions}

The stability of anisotropic stellar structures is an important
topic of discussion in the background of astrophysical world. One
such method to estimate the stable nature of quark compositions is
regarded as adiabatic index of the form
\begin{equation}\label{59b}
\Gamma_{r}=\frac{\mathbb{U}+\mathbb{P}_{r}}{\mathbb{P}_{r}}\bigg(\frac{d\mathbb{P}_{r}}{d\mathbb{U}}\bigg),\quad
\Gamma_{t}=\frac{\mathbb{U}+\mathbb{P}_{t}}{\mathbb{P}_{t}}\bigg(\frac{d\mathbb{P}_{t}}{d\mathbb{U}}\bigg),
\end{equation}
where $\Gamma_{r}$ and $\Gamma_{t}$ are its radial and temporal
components, whose values must be greater than $\frac{4}{3}$ to
depict the stable distributions \cite{25a}. In addition to this, the
causality condition is an alternative approach according to which
the speed of light exceeds the speed of sound. The radial and
transverse velocities, respectively, are two components described by
$\mathbb{V}^{2}_{r}$ and $\mathbb{V}^{2}_{t}$. For the stable
configuration, these constituents must fulfill the conditions
$0<\mathbb{V}^{2}_{r}< 1$ and $0<\mathbb{V}^{2}_{t}<1$, and they are
mathematically expressed as
\begin{figure}\center
\epsfig{file=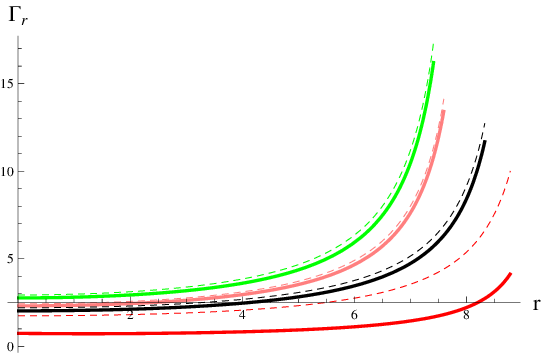,width=0.4\linewidth}\epsfig{file=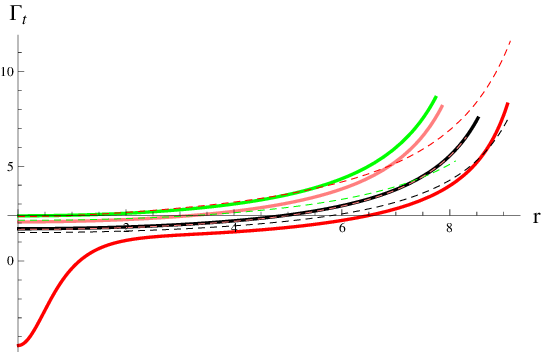,width=0.4\linewidth}
\epsfig{file=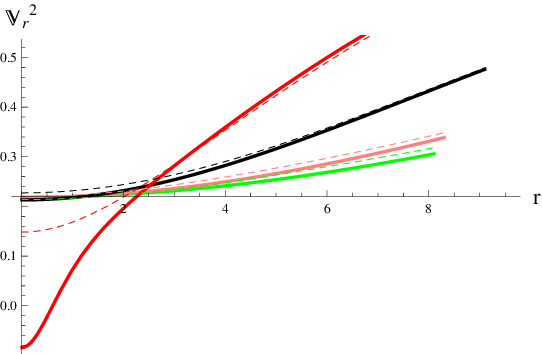,width=0.4\linewidth}\epsfig{file=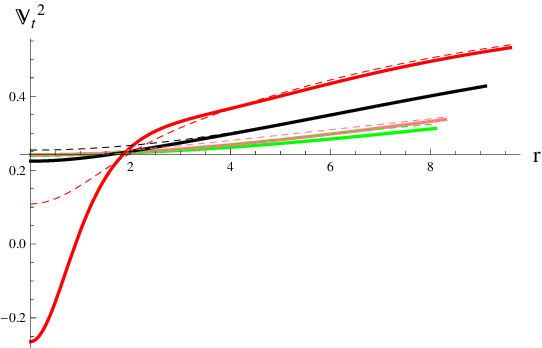,width=0.4\linewidth}
\caption{Behavior of adiabatic index and causality condition versus
$r$ for Her X-I, LMC X-4, 4U1820-30 and PSR J1614-2230 quark
candidates.}
\end{figure}
\begin{equation}\label{59a}
\mathbb{V}^{2}_{r}=\frac{d\mathbb{P}_{r}}{d\mathbb{U}},\quad\quad
\mathbb{V}^{2}_{t}=\frac{d\mathbb{P}_{t}}{d\mathbb{U}}.
\end{equation}
In other words, we can also say that both the aforementioned
components must lie between the range of 0 and 1 \cite{24ab}. Figure
\textbf{7} exhibits both criteria, from which we can notice that all
the stellar stars meet the necessary requisites for
$\mathcal{Q}=0.1$. For the higher charge value, the candidates Her
X-I, 4U1820-30 and LMC X-4 portray the stable structures whereas PSR
J1614-2230 becomes inconsistent through both techniques as not
satisfying the required limits (Figure \textbf{7}).

\section{Conclusion}

The main motivation of this paper is to study the role of
electromagnetic field on the substantial features of Her X-I, PSR
J1614-2230, 4U1820-30 and LMC X-4. In this regard, we have
considered the static spherical compositions occupying anisotropic
source in their interior. The modified field equations are
formulated using the quadratic model $f(G,T)=G^2+\varrho T$ of
$f(G,T)$ gravity. The radial pressure and density are interlinked
with bag constant in the MIT bag model EoS which transforms the
modified field equations in terms of $\mathfrak{B}_c$. The radial
and temporal metric potentials of Tolman IV solution have been used
to find the anisotropic solution with $\mathfrak{X}^2$,
$\mathfrak{C}^2$ and $\mathfrak{Z}^2$ as unknowns. The interior
spacetime is matched with the Reissner-Nordstr$\ddot{o}$m metric
(chosen as outer geometry) to compute these unknown variables.
Further, the boundary conditions represent these constants into
masses and radii of the proposed quark stars. The expression of
$\mathbb{P}_r$ in \textbf{Appendix A} yields the bag constant
equation through which we have evaluated the value of
$\mathfrak{B}_{c}$ for each strange star corresponding to
$\mathcal{Q}=0.1$ and 1.5 (Tables \textbf{2} and \textbf{3}) by
using their observational data. The graphical analysis of all the
considered compact bodies have been discussed for two values of
charge.

The physical entities (energy density and pressure components) have
shown the acceptable behavior, i.e., the maximal values at the
center which decrease gradually with increasing $r$. The
compatibility of the solution has been confirmed through the
behavior of metric potentials (Figure \textbf{1}). The viability of
all quark candidates has been investigated through energy conditions
that ensure the occurrence of regular matter within these stars
corresponding to both charges. It is also found that compactness,
mass and redshift factors satisfy their necessary ranges. The
stellar compact objects Her X-I, LMC X-4 and 4U1820-30 depict the
stable nature for both charges as they meet the confined bounds. The
star candidate PSR J1614-2230 portrays a stable structure when
$\mathcal{Q}=0.1$, but becomes unstable for $\mathcal{Q}=1.5$. In GR
\cite{30}, the stars LMC X-4 and Her X-I were found to be more
denser in comparison to this present work. However, one can also
notice that more appropriate results are established for 4U1820-30
as compared to \cite{31}. Sharif et al \cite{31a} established a
viable and stable charged 4U 1608-52 star by assuming two particular
values of $\mathfrak{B}_c$ in $f(G,T)$ gravity. It is found that the
proposed solutions reduce to GR when $\varrho=\varsigma=0$ is
imposed on the model \eqref{60}.

\section*{Appendix A}

The physical variables corresponding to unknowns involved in Tolman
IV solution are
\begin{align}\nonumber
\mathbb{U}&=\frac{1}{2 \mathfrak{Z}^8 (\varrho +8 \pi )
\big(\mathfrak{X}^2+2 r^2\big)^7}\{\mathfrak{X}^6 \mathfrak{Z}^2
\big(6 \mathfrak{X}^6 \big(\mathfrak{Z}^6-384\big)-13632
\mathfrak{X}^4 \mathfrak{Z}^2\\\nonumber&-26688 \mathfrak{X}^2
\mathfrak{Z}^4+\mathfrak{X}^8 \mathfrak{Z}^4 \big(2 \mathfrak{B}_{c}
\mathfrak{Z}^2 (\varrho +8 \pi )+3\big)-17280
\mathfrak{Z}^6\big)+256 \mathfrak{B}_{c} \mathfrak{Z}^8 (\varrho
\\\nonumber&+8 \pi ) r^{14}+32 r^{12} \big(\mathfrak{X}^2
\big(28 \mathfrak{B}_{c} \mathfrak{Z}^8 (\varrho +8 \pi )+3
\mathfrak{Z}^6-72\big)+6 \mathfrak{Z}^2
\big(\mathfrak{Z}^6-24\big)\big)\\\nonumber&+48 r^{10} \big(2
\mathfrak{X}^2 \mathfrak{Z}^2 \big(7
\mathfrak{Z}^6-8\big)+\mathfrak{X}^4 \big(28 \mathfrak{B}_{c}
\mathfrak{Z}^8 (\varrho +8 \pi )+7 \mathfrak{Z}^6-32\big)+96
\mathfrak{Z}^4\big)\\\nonumber&+32 r^8 \big(6 \mathfrak{X}^4
\mathfrak{Z}^2 \big(5 \mathfrak{Z}^6-164\big)-1056 \mathfrak{X}^2
\mathfrak{Z}^4+\mathfrak{X}^6 \big(35 \mathfrak{B}_{c}
\mathfrak{Z}^8 (\varrho +8 \pi )+15
\mathfrak{Z}^6\\\nonumber&-216\big)+96 \mathfrak{Z}^6\big)+8 r^6
\big(\mathfrak{X}^6 \big(90 \mathfrak{Z}^8-8256
\mathfrak{Z}^2\big)-7776 \mathfrak{X}^4 \mathfrak{Z}^4+2112
\mathfrak{X}^2 \mathfrak{Z}^6\\\nonumber&+5 \mathfrak{X}^8 \big(14
\mathfrak{B}_{c} \mathfrak{Z}^8 (\varrho +8 \pi )+9
\mathfrak{Z}^6-384\big)\big)+6 \mathfrak{X}^2 r^4
\big(\mathfrak{X}^6 \big(50 \mathfrak{Z}^8-512
\mathfrak{Z}^2\big)\\\nonumber&+10208 \mathfrak{X}^4
\mathfrak{Z}^4+19776 \mathfrak{X}^2 \mathfrak{Z}^6+\mathfrak{X}^8
\big(28 \mathfrak{B}_{c} \mathfrak{Z}^8 (\varrho +8 \pi )+25
\mathfrak{Z}^6-768\big)\\\nonumber&+6912
\mathfrak{Z}^8\big)+\mathfrak{X}^4 r^2 \big(66 \mathfrak{X}^6
\mathfrak{Z}^2 \big(\mathfrak{Z}^6+384\big)+71424 \mathfrak{X}^4
\mathfrak{Z}^4+78144 \mathfrak{X}^2
\mathfrak{Z}^6\\\nonumber&+\mathfrak{X}^8 \big(28 \mathfrak{B}_{c}
\mathfrak{Z}^8 (\varrho +8 \pi )+33 \mathfrak{Z}^6+3072\big)+24192
\mathfrak{Z}^8\big)\}^{-1},
\\\nonumber \mathbb{P}_{r}&=-\frac{1}{2
\mathfrak{Z}^8 (\varrho +8 \pi ) \big(\mathfrak{X}^2+2
r^2\big)^7}\{\mathfrak{X}^6 \mathfrak{Z}^2 \big(-2 \mathfrak{X}^6
\big(\mathfrak{Z}^6-384\big)+4544 \mathfrak{X}^4
\mathfrak{Z}^2\\\nonumber&+8896 \mathfrak{X}^2
\mathfrak{Z}^4+\mathfrak{X}^8 \mathfrak{Z}^4 \big(2 \mathfrak{B}_{c}
\mathfrak{Z}^2 (\varrho +8 \pi )-1\big)+5760 \mathfrak{Z}^6\big)+256
\mathfrak{B}_{c} \mathfrak{Z}^8 (\varrho \\\nonumber&+8 \pi )
r^{14}+32 r^{12} \big(\mathfrak{X}^2 \big(28 \mathfrak{B}_{c}
\mathfrak{Z}^8 (\varrho +8 \pi )-\mathfrak{Z}^6+24\big)-2
\mathfrak{Z}^2 \big(\mathfrak{Z}^6-24\big)\big)\\\nonumber&+16
r^{10} \big(-2 \mathfrak{X}^2 \mathfrak{Z}^2 \big(7
\mathfrak{Z}^6-8\big)+\mathfrak{X}^4 \big(84 \mathfrak{B}_{c}
\mathfrak{Z}^8 (\varrho +8 \pi )-7 \mathfrak{Z}^6+32\big)-96
\mathfrak{Z}^4\big)\\\nonumber&+32 r^8 \big(\mathfrak{X}^4 \big(328
\mathfrak{Z}^2-10 \mathfrak{Z}^8\big)+352 \mathfrak{X}^2
\mathfrak{Z}^4+\mathfrak{X}^6 \big(35 \mathfrak{B}_{c}
\mathfrak{Z}^8 (\varrho +8 \pi )-5
\mathfrak{Z}^6+72\big)\\\nonumber&-32 \mathfrak{Z}^6\big)+8 r^6
\big(\mathfrak{X}^6 \big(2752 \mathfrak{Z}^2-30
\mathfrak{Z}^8\big)+2592 \mathfrak{X}^4 \mathfrak{Z}^4-704
\mathfrak{X}^2 \mathfrak{Z}^6+5 \mathfrak{X}^8 \big(14
\mathfrak{B}_{c} \mathfrak{Z}^8 \\\nonumber&\times(\varrho +8 \pi
)-3 \mathfrak{Z}^6+128\big)\big)+2 \mathfrak{X}^2 r^4
\big(\mathfrak{X}^6 \big(512 \mathfrak{Z}^2-50
\mathfrak{Z}^8\big)-10208 \mathfrak{X}^4
\mathfrak{Z}^4\\\nonumber&-19776 \mathfrak{X}^2
\mathfrak{Z}^6+\mathfrak{X}^8 \big(84 \mathfrak{B}_{c}
\mathfrak{Z}^8 (\varrho +8 \pi )-25 \mathfrak{Z}^6+768\big)-6912
\mathfrak{Z}^8\big)-\mathfrak{X}^4 r^2 \\\nonumber&\times\big(22
\mathfrak{X}^6 \mathfrak{Z}^2 \big(\mathfrak{Z}^6+384\big)+23808
\mathfrak{X}^4 \mathfrak{Z}^4+26048 \mathfrak{X}^2
\mathfrak{Z}^6+\mathfrak{X}^8 \big(-28 \mathfrak{B}_{c}
\mathfrak{Z}^8 (\varrho +8 \pi )\\\nonumber&+11
\mathfrak{Z}^6+1024\big)+8064 \mathfrak{Z}^8\big)\}^{-1},
\\\nonumber \mathbb{P}_{t}&=\frac{1}{4
r^4 \big(2 r^2+\mathfrak{X}^2\big)^7 \mathfrak{Z}^8 (\varrho +6 \pi
) (\varrho +8 \pi )}\{8 \pi  \big(64 \big(4 \mathfrak{B}_{c} \varrho
\mathfrak{Z}^8-9 \mathfrak{Z}^6+108\big) r^{18}\\\nonumber&+64
\big(2 \big(7 \mathfrak{B}_{c} \varrho \mathfrak{Z}^8-15
\mathfrak{Z}^6+171\big) \mathfrak{X}^2+3 \mathfrak{Z}^2
\big(\mathfrak{Z}^6-24\big)\big) r^{16}+48 \big(\big(28
\mathfrak{B}_{c} \varrho \mathfrak{Z}^8\\\nonumber&-57
\mathfrak{Z}^6+528\big) \mathfrak{X}^4+4 \mathfrak{Z}^2 \big(3
\mathfrak{Z}^6-136\big) \mathfrak{X}^2-8 \mathfrak{Z}^4 \big(q^2
\mathfrak{Z}^4+22\big)\big) r^{14}+16 \big(\big(70
\mathfrak{B}_{c}\\\nonumber&\times \varrho \mathfrak{Z}^8-135
\mathfrak{Z}^6+1776\big) \mathfrak{X}^6+15 \mathfrak{Z}^2 \big(3
\mathfrak{Z}^6+104\big) \mathfrak{X}^4-84 \mathfrak{Z}^4 \big(q^2
\mathfrak{Z}^4-46\big) \mathfrak{X}^2\\\nonumber& +576
\mathfrak{Z}^6\big) r^{12}+4 \big(\big(140 \mathfrak{B}_{c} \varrho
\mathfrak{Z}^8-255 \mathfrak{Z}^6+4224\big) \mathfrak{X}^8+24
\mathfrak{Z}^2 \big(5 \mathfrak{Z}^6-72\big)
\mathfrak{X}^6\\\nonumber&-24 \mathfrak{Z}^4 \big(21 q^2
\mathfrak{Z}^4+982\big) \mathfrak{X}^4-36672 \mathfrak{Z}^6
\mathfrak{X}^2\big) r^{10}+12 \mathfrak{X}^2 \big(2 \big(7
\mathfrak{B}_{c} \varrho \mathfrak{Z}^8-12
\mathfrak{Z}^6\\\nonumber&-256\big) \mathfrak{X}^8+\mathfrak{Z}^2
\big(15 \mathfrak{Z}^6-5984\big) \mathfrak{X}^6-28 \mathfrak{Z}^4
\big(5 q^2 \mathfrak{Z}^4+442\big) \mathfrak{X}^4-3904
\mathfrak{Z}^6 \mathfrak{X}^2\\\nonumber&+8064 \mathfrak{Z}^8\big)
r^8+\mathfrak{X}^4 \mathfrak{Z}^2 \big(\mathfrak{Z}^4 \big(28
\mathfrak{B}_{c} \mathfrak{Z}^2 \varrho -45\big) \mathfrak{X}^8+12
\big(3 \mathfrak{Z}^6+704\big) \mathfrak{X}^6-840
\mathfrak{Z}^2\\\nonumber&\times \big(q^2 \mathfrak{Z}^4-80\big)
\mathfrak{X}^4+143808 \mathfrak{Z}^4 \mathfrak{X}^2+98496
\mathfrak{Z}^6\big) r^6 +\mathfrak{X}^6 \mathfrak{Z}^4
\big(\mathfrak{Z}^2 \big(2 \mathfrak{B}_{c} \mathfrak{Z}^2 \varrho
-3\big) \mathfrak{X}^8\\\nonumber&+3 \mathfrak{Z}^4
\mathfrak{X}^6-12 \big(21 q^2 \mathfrak{Z}^4+184\big)
\mathfrak{X}^4-11328 \mathfrak{Z}^2 \mathfrak{X}^2-12960
\mathfrak{Z}^4\big) r^4-42 q^2 \mathfrak{X}^{12} \mathfrak{Z}^8
r^2\\\nonumber&-3 q^2 \mathfrak{X}^{14} \mathfrak{Z}^8\big)+\varrho
\big(64 \big(4 \mathfrak{B}_{c} \varrho \mathfrak{Z}^8-9
\mathfrak{Z}^6+108\big) r^{18}+64 \big(\big(14 \mathfrak{B}_{c}
\varrho \mathfrak{Z}^8-29 \mathfrak{Z}^6\\\nonumber&+318\big)
\mathfrak{X}^2+5 \mathfrak{Z}^2 \big(\mathfrak{Z}^6-24\big)\big)
r^{16}+16 \big(\big(84 \mathfrak{B}_{c} \varrho \mathfrak{Z}^8-157
\mathfrak{Z}^6+1520\big) \mathfrak{X}^4\\\nonumber&+64
\mathfrak{Z}^2 \big(\mathfrak{Z}^6-26\big) \mathfrak{X}^2-24
\mathfrak{Z}^4 \big(q^2 \mathfrak{Z}^4+14\big)\big) r^{14}+16
\big(\big(70 \mathfrak{B}_{c} \varrho \mathfrak{Z}^8-115
\mathfrak{Z}^6\\\nonumber&+1488\big) \mathfrak{X}^6+\mathfrak{Z}^2
\big(85 \mathfrak{Z}^6+248\big) \mathfrak{X}^4+\big(2456
\mathfrak{Z}^4-84 q^2 \mathfrak{Z}^8\big) \mathfrak{X}^2+704
\mathfrak{Z}^6\big) r^{12}\\\nonumber&+4 \big(\big(1664+140
\mathfrak{B}_{c} \varrho \mathfrak{Z}^8-195 \mathfrak{Z}^6\big)
\mathfrak{X}^8+16 \mathfrak{Z}^2 \big(15 \mathfrak{Z}^6-796\big)
\mathfrak{X}^6-168 \mathfrak{Z}^4\\\nonumber&\times \big(3 q^2
\mathfrak{Z}^4+202\big) \mathfrak{X}^4-33856 \mathfrak{Z}^6
\mathfrak{X}^2\big) r^{10}+4 \mathfrak{X}^2 \big(\big(42
\mathfrak{B}_{c} \varrho \mathfrak{Z}^8-2304-47 \mathfrak{Z}^6\big)
\\\nonumber&\times\mathfrak{X}^8+\mathfrak{Z}^2 \big(95 \mathfrak{Z}^6-18464\big)
\mathfrak{X}^6-20 \mathfrak{Z}^4 \big(21 q^2
\mathfrak{Z}^4+1346\big) \mathfrak{X}^4+8064 \mathfrak{Z}^6
\mathfrak{X}^2\\\nonumber&+31104 \mathfrak{Z}^8\big)
r^8+\mathfrak{X}^4 \big(\big(28 \mathfrak{B}_{c} \varrho
\mathfrak{Z}^8-23 \mathfrak{Z}^6+2048\big) \mathfrak{X}^8+16
\mathfrak{Z}^2 \big(5 \mathfrak{Z}^6+1584\big)
\mathfrak{X}^6\\\nonumber&-24 \mathfrak{Z}^4 \big(35 q^2
\mathfrak{Z}^4-4784\big) \mathfrak{X}^4+195904 \mathfrak{Z}^6
\mathfrak{X}^2+114624 \mathfrak{Z}^8\big) r^6+\mathfrak{X}^6
\mathfrak{Z}^2 \big(\mathfrak{Z}^4 \\\nonumber&\times\big(2
\mathfrak{B}_{c} \mathfrak{Z}^2 \varrho -1\big)
\mathfrak{X}^8+\big(7 \mathfrak{Z}^6-1536\big) \mathfrak{X}^6-4
\mathfrak{Z}^2 \big(63 q^2 \mathfrak{Z}^4+2824\big)
\mathfrak{X}^4\\\nonumber&-29120 \mathfrak{Z}^4 \mathfrak{X}^2-24480
\mathfrak{Z}^6\big) r^4-42 q^2 \mathfrak{X}^{12} \mathfrak{Z}^8
r^2-3 q^2 \mathfrak{X}^{14} \mathfrak{Z}^8\big)\}^{-1}.
\end{align}


\begin{thebibliography}{40}

\bibitem{1} W. Baade, F. Zwicky, Remarks on super-novae and cosmic rays, Phys. Rev. 46 (1934)
76, https://doi.org/10.1103/Phys. Rev.46.76.2.

\bibitem{2} E. Witten, Cosmic separation of phases, Phys. Rev. D 30 (1984) 272, https://doi.10.1103/physrevd.30.272; X.D. Li, Z.G. Dai, Z.R.
Wang, Is Her X-1 a strange star?, Astron. Astrophys. 303 (1995) L1;
I. Bombaci, Observational evidence for strange matter in compact
objects from the x-ray burster 4U 1820-30, Phys. Rev. C 55 (1997)
1587, https://doi.org/10.1103/Phys Rev C.55.1587.

\bibitem{3} M. Ruderman, Pulsars: structure and dynamics, Annu. Rev. Astron. Astrophys. 10 (1972) 427, https://doi.org/10.1146/annurev.aa.10.090172.002235.

\bibitem{5} K. Dev,  M. Gleiser, Anisotropic stars: exact solutions, Gen. Relativ. Grav. 34 (2002) 1793, 0001-7701/02/1100-1793/0.

\bibitem{6} S.K.M. Hossein,  et al., Anisotropic compact stars with variable cosmological constant,  Int. J. Mod. Phys. D 21 (2012) 1250088,
https://doi.org/10.1142/S0218271812500885.

\bibitem{7} M. Kalam, et al., Anisotropic quintessence stars, Astrophys. Space Sci. 349 (2014) 865, http://doi.10.1007/s10509-013-1677-x.

\bibitem{8} S.K. Maurya, et al., Generalised model for anisotropic compact stars, Eur. Phys. J. C 76 (2016) 693, http://doi 10.1140/epjc/s10052-016-4527-5.

\bibitem{8a} M.H. Murad, Some analytical models of anisotropic strange stars, Astrophys. Space Sci. 361 (2016) 20, http://doi 10.1007/s10509-015-2582-2.

\bibitem{8b} K.D. Matondo, S.D. Maharaj, S. Ray, Charged isotropic model with conformal symmetry, Astrophys. Space Sci. 363 (2018) 187,
https://doi.org/10.1007/s10509-018-3410-2.

\bibitem{8c} P. Mafa Takisa, S.D. Maharaj, L.L. Leeuw, Effect of electric charge on conformal compact stars, Eur. Phys. J. C
79 (2019) 8, https://doi.org/10.1140/epjc/s10052-018-6519-0; S.K.
Maurya, F. Tello-Ortiz, Charged anisotropic compact star in $f(R,T)$
gravity: A minimal geometric deformation gravitational decoupling
approach, Phys. Dark Universe 27 (2020) 100442,
https://doi.org/10.1016/j.dark.2019.100442; G. Panotopoulos,  et
al., Charged polytropic compact stars in 4D Einstein Gauss-Bonnet
gravity, Chin. J. Phys. 77 (2022) 2106,
https://doi.org/10.1016/j.cjph.2022.01.008; M. Sharif, T. Naseer,
Influence of charge on extended decoupled anisotropic solutions in
$f(R,T,R_{\lambda\xi}T^{\lambda\xi})$ gravity, Indian J. Phys. 96
(2022) 4373, https://doi.org/10.1007/s12648-022-02339-7.

\bibitem{9} P. Haensel, J.L. Zdunik, R. Schaefer, Strange quark stars, Astron. Astrophys. 160 (1986) 121; F. Weber, Strange quark matter and compact stars, Prog.
Part. Nucl. Phys. 54 (2005) 193,
https://doi.org/10.1016/j.ppnp.2004.07.001; M.A. Perez Garcia, J.
Silk, J.R. Stone, Dark matter, neutron stars, and strange quark
matter, Phys. Rev. Lett. 105 (2010) 141101, http://doi:
10.1103/PhysRevLett.105.141101; H. Rodrigues, S.B. Duarte, J.C.T. De
Oliveira, Massive compact stars as quark stars, Astrophys. J. 730
(2011) 31, http://doi:10.1088/0004-637X/730/1/31.

\bibitem{9a} P.B. Demorest, et al., A two-solar-mass neutron star measured using Shapiro delay, Nature 467 (2010) 1081, http://doi:10.1038/nature09466.

\bibitem{10} F. Rahaman, et al., A new deterministic model of strange stars, Eur. Phys. J. C 74 (2014) 3126, http://doi 10.1140/epjc/s10052-014-3126-6.

\bibitem{11} P. Bhar, A new hybrid star model in Krori-Barua spacetime, Astrophys. Space Sci. 357 (2015) 46, http://doi 10.1007/s10509-015-2271-1.

\bibitem{12} D. Deb, et al., Relativistic model for anisotropic strange stars, Ann. Phys. 387 (2017) 239, https://doi.org/10.1016/j.aop.2017.10.010;
D. Deb, et al., Anisotropic strange stars in the Einstein-Maxwell
spacetime, Eur. Phys. J. C 78 (2018) 465,
https://doi.org/10.1140/epjc/s10052-018-5930-x.

\bibitem{13} M. Sharif, A. Waseem, Anisotropic quark stars in $f(R,T)$ gravity, Eur. Phys. J. C 78 (2018) 868,
https://doi.org/10.1140/epjc/s10052-018-6363-2; Role of
curvature-matter coupling on anisotropic strange stars, Chin. J.
Phys. 63 (2020) 92, https://doi.org/10.1016/j.cjph.2019.11.006; M.
Sharif, A. Majid, Compact stars with MIT bag model in massive
Brans-Dicke gravity, Astrophys. Space Sci. 366 (2021) 54,
https://doi.org/10.1007/s10509-021-03962-2; A. Majid, M. Sharif,
Quark stars in massive Brans Dicke gravity with Tolman Kuchowicz
spacetime, Universe 6 (2020) 124,
https://doi:10.3390/universe6080124; M. Sharif, T. Naseer, Effects
of non-minimal matter-geometry coupling on embedding class-one
anisotropic solutions, Phys. Scr. 97 (2022) 055004, https://doi
10.1088/1402-4896/ac5ed4; Study of charged compact stars in
non-minimally coupled gravity, Fortschr. Phys. 71 (2022) 2200147,
https://doi.org/10.1002/prop.202200147; Study of anisotropic compact
stars in $f(R,T,R_{\chi\xi}T^{\chi\xi\xi})$ gravity, Pramana 96
(2022) 119, https://doi.org/10.1007/s12043-022-02357-4.

\bibitem{32} S. Capozziello, A. Stabile, A. Troisi, Spherical symmetry in $f(R)$-gravity
Class. Quantum Grav. 25 (2008) 085004, https://doi
10.1088/0264-9381/25/8/085004; S. Capozziello, E. De Filippis, V.
Salzano, Modelling clusters of galaxies by $f(R)$ gravity, Mon. Not.
R. Astron. Soc. 394 (2009) 947,
https://doi.org/10.1111/j.1365-2966.2008.14382.x; S. Nojiri, S.D.
Odintsov, Unified cosmic history in modified gravity: from $f(R)$
theory to Lorentz non-invariant models, Phys. Rep. 505 (2011) 59,
https://doi.org/10.1016/j.physrep.2011.04.001.

\bibitem{33} M. Sharif, H.R. Kausar, Effects of $f(R)$ model on the dynamical instability of expansion free gravitational collapse,
 J. Cosmol. Astropart. Phys. 07 (2011) 022, https://doi:10.1088/1475-7516/2011/07/022;
A.V. Astashenok, S. Capozziello, S.D. Odintsov, Extreme neutron
stars from extended theories of gravity, J. Cosmol. Astropart. Phys.
2015 (2015) 001, https://doi:10.1088/1475-7516/2015/01/001;
Nonperturbative models of quark stars in $f(R)$ gravity, Phys. Lett.
B 742 (2015) 160, https://doi.org/10.1016/j.physletb.2015.01.030.

\bibitem{34} T. Harko, et al., $f(R,T)$  gravity, Phys. Rev. D 84 (2011) 024020, https://doi.10.1103/PhysRevD.84.024020.

\bibitem{35} M. Sharif, M. Zubair, Thermodynamic behavior of particular $f(R,T)$ gravity
models, J. Exp. Theor. Phys. 117 (2013) 248,
https://doi.10.1134/S1063776113100075; H. Shabani, M. Farhoudi,
$f(R,T)$ cosmological models in phase space, Phys. Rev. D 88 (2013)
044048, https://doi.org/10.1103/PhysRevD.88.044048; A. Alhamzawi, R.
Alhamzawi, Gravitational lensing by $f(R,T)$ gravity, Int. J. Mod.
Phys. D 25 (2016) 1650020,
https://doi.org/10.1142/S0218271816500206; M. Sharif, A. Siddiqa,
Study of charged stellar structures in $f(R,T)$ gravity, Eur. Phys.
J. Plus 132 (2017) 529, https://doi.10.1140/epjp/i2017-11810-4.

\bibitem{36} S. Nojiri, S.D. Odintsov, Modified Gauss-Bonnet theory as gravitational alternative for dark energy, Phys. Lett. B 631 (2005) 1,
https://doi.org/10.1016/j.physletb.2005.10.010.

\bibitem{14} K. Bamba, et al., Finite-time future singularities in modified Gauss-Bonnet and $f(R,G)$ gravity and singularity avoidance,
Eur. Phys. J. C 67 (2010) 295,
https://doi.10.1140/epjc/s10052-010-1292-8.

\bibitem{15} R. Myrzakulov, D. Saez-Gomez, A. Tureanu, On the $\Lambda$CDM Universe in $f(G)$ gravity, Gen. Relativ. Grav.
43 (2011) 1671, https://doi.10.1007/s10714-011-1149-y.

\bibitem{17} M. Sharif, H. Ismat Fatima, Noether symmetries in $f(G)$ gravity, J. Exp. Theor. Phys. 122 (2016) 104, https://doi.10.1134/S1063776116010192.

\bibitem{16} K. Bamba, et al., Energy conditions in modified $f(G)$ gravity, Gen. Relativ. Grav. 49 (2017) 112, https://doi.10.1007/s10714-017-2276-x.

\bibitem{18} M. Sharif, A. Ramzan, Anisotropic compact stellar objects in modified Gauss-Bonnet gravity, Phys. Dark Universe 30 (2020) 100737,
https://doi.org/10.1016/j.dark.2020.100737.

\bibitem{37} M. Sharif, A. Ikram, Energy conditions in $f(G,T)$ gravity, Eur. Phys. J. C 76 (2016) 640, https://doi.10.1140/epjc/s10052-016-4502-1.

\bibitem{19} M.F. Shamir, M. Ahmad, Emerging anisotropic compact stars in $f(G,T)$ gravity, Eur. Phys. J. C 77 (2017) 674,
https://doi.10.1140/epjc/s10052-017-5239-1.

\bibitem{20} S.K. Maurya, et al.,  Anisotropic stars in  $f(G,T)$ gravity under class I space-time, Eur. Phys. J. Plus 135 (2020) 824,
https://doi.org/10.1140/epjp/s13360-020-00832-8.

\bibitem{21} M. Sharif, A. Naeem, Anisotropic solution for compact objects in $f(G,T)$ gravity, Int. J. Mod. Phys. A 35 (2020) 2050121,
https://doi.org/10.1142/S0217751X20501213.

\bibitem{21a} M. Sharif, K. Hassan, Electromagnetic effects on the complexity of static cylindrical object in $f(G,T)$ gravity, Eur. Phys. J. Plus
137 (2022) 1380, https://doi.org/10.1140/epjp/s13360-022-03612-8;
Complexity factor for static cylindrical objects in $f(G,T)$
gravity, Pramana 96 (2022) 50,
https://doi.org/10.1007/s12043-022-02298-y; Complexity of dynamical
cylindrical system in $f(G,T)$ gravity, Mod. Phys. Lett. A 37 (2022)
2250027, https://doi.org/10.1142/S0217732322500274; Analysis of
complexity factor for charged dissipative configuration in modified
gravity, Eur. Phys. J. Plus 138 (2023) 787,
https://doi.org/10.1140/epjp/s13360-023-04417-z; Complexity for
dynamical anisotropic sphere in $f(G,T)$ gravity, Chin. J. Phys 77
(2022) 1479, https://doi.org/10.1016/j.cjph.2021.11.038; Complexity
of charged dynamical spherical system in modified gravity, Chin. J.
Phys 84 (2023) 152, https://doi.org/10.1016/j.cjph.2023.03.024.

\bibitem{25} M. Kalma, et al., A relativistic model for strange quark star, Int. J. Theor. Phys. 52 (2013) 3319, https://doi.10.1007/s10773-013-1629-9;
J.D.V. Arbanil, M. Malheiro, Radial stability of anisotropic strange
quark stars, J. Cosmol. Astropart. Phys. 2016 (2016) 012,
doi:10.1088/1475-7516/2016/11/012; M. Sharif, A. Majid, Anisotropic
strange stars through embedding technique in massive Brans-Dicke
gravity, Eur. Phys. J. C 135 (2020) 558,
https://doi.org/10.1140/epjp/s13360-020-00574-7.

\bibitem{22c} H.A. Buchdahl, General relativistic fluid spheres, Phys. Rev. 116 (1959) 1027, https://doi.org/10.1103/PhysRev.116.1027.

\bibitem{26} E. Farhi, R.L. Jaffe, Strange matter, Phys. Rev. D 30 (1984) 2379, https://doi.org/10.1103/PhysRevD.30.2379.

\bibitem{27} N. Stergioulas, Rotating stars in relativity, Living Rev. Relativ. 6 (2003) 3, https://doi:10.12942/lrr-2003-3.

\bibitem{28} R.X. Xu, What can the redshift observed in EXO 0748-676 tell us?, Chin. J. Astron. Astrophys. 3 (2003) 33, https://doi:10.1088/1009-9271/3/1/33.

\bibitem{29} F. Rahman, et al. Strange stars in Krori-Barua spacetime, Eur. Phys. J. C 72 (2012) 2071,
https://doi:10.1140/epjc/s10052-012-2071-5.

\bibitem{29a} N.K. Glendenning, Compact stars:
Nuclear physics, particle physics and general relativity (Springer
Science Business Media, 2012),
https://doi.org/10.1007/978-1-4684-0491-3; J.D.V. Arbanil, J.P.S.
Lemos, V.T. Zanchin, Incompressible relativistic spheres:
Electrically charged stars, compactness bounds, and quasiblack hole
configurations, Phys. Rev. D 89 (2014) 104054,
https://doi.org/10.1103/PhysRevD.89.104054; J.P.S. Lemos, et al.,
Compact stars with a small electric charge: the limiting radius to
mass relation and the maximum mass for incompressible matter, Eur.
Phys. J. C 75 (2015) 76, https://doi.10.1140/epjc/s10052-015-3274-3;
J.D.V. Arbanil, V.T. Zanchin, Relativistic polytropic spheres with
electric charge: Compact stars, compactness and mass bounds, and
quasiblack hole configurations, Phys. Rev. D 97 (2018) 104045,
https://doi:10.1103/PhysRevD.97.104045.

\bibitem{25a} H. Heintzmann, W. Hillebrandt, Neutron stars with an anisotropic equation of state-mass, redshift and stability, Astron. Astrophys.
38 (1975) 51.

\bibitem{24ab} H. Abreu, H. Hernandez, L.A. Nunez, Sound speeds, cracking and the stability of self-gravitating anisotropic compact objects,
Class. Quantum Grav. 24 (2007) 4631,
https://doi:10.1088/0264-9381/24/18/005.

\bibitem{30} P.C. Fulara, A. Sah, A spherical relativistic anisotropic compact star model, Int. J. Astron. Astrophys. 8 (2018) 46,
https://doi:10.4236/ijaa.2018.81004 .

\bibitem{31} M.F. Shamir, M. Ahmad, Emerging anisotropic compact stars in $f(G,T)$ gravity, Eur. Phys. J. C 77 (2017)
674, https://doi:10.1140/epjc/s10052-017-5239-1.

\bibitem{31a} M. Sharif, A. Naeem, A. Ramzan, Charged anisotropic strange stars in $f(G,T)$ gravity, Astrophys. Space Sci. 367 (2022) 21,
https://doi.org/10.1007/s10509-022-04052-7.

\end{thebibliography}
\end{document}